\documentclass[iop]{emulateapj}
\usepackage{nicefrac}
\usepackage[colorlinks=true, urlcolor=blue]{hyperref}

\newcommand{\myemail}{\href{mailto:iarcavi@lcogt.net}{iarcavi@lcogt.net}}
\def\gtorder{\mathrel{\raise.3ex\hbox{$>$}\mkern-14mu
             \lower0.6ex\hbox{$\sim$}}} 
\def\ltorder{\mathrel{\raise.3ex\hbox{$<$}\mkern-14mu
             \lower0.6ex\hbox{$\sim$}}}
\def\ltsima{$\; \buildrel < \over \sim \;$}
\def\simlt{\lower.5ex\hbox{\ltsima}}
\def\gtsima{$\; \buildrel > \over \sim \;$}
\def\simgt{\lower.5ex\hbox{\gtsima}} 

\shorttitle{H- and He-rich TDE Candidates}
\shortauthors{Arcavi et al.}

\begin{document} 


\title{A Continuum of H- to He-Rich Tidal Disruption Candidates With a Preference for E+A Galaxies}


\author{Iair~Arcavi\altaffilmark{1,2,3},
Avishay~Gal-Yam\altaffilmark{1},
Mark~Sullivan\altaffilmark{4},
Yen-Chen~Pan\altaffilmark{5},
S.~Bradley~Cenko\altaffilmark{6,7},
Assaf~Horesh\altaffilmark{1},
Eran~O.~Ofek\altaffilmark{1},
Annalisa~De~Cia\altaffilmark{1},
Lin~Yan\altaffilmark{8},
Chen-Wei~Yang\altaffilmark{8,9},
D.~A.~Howell\altaffilmark{2,10},
David~Tal\altaffilmark{1},
Shrinivas~R.~Kulkarni\altaffilmark{11},
Shriharsh~P.~Tendulkar\altaffilmark{11},
Sumin~Tang\altaffilmark{11,3},
Dong~Xu\altaffilmark{12},
Assaf~Sternberg\altaffilmark{13,14},
Judith~G.~Cohen\altaffilmark{11},
Joshua~S.~Bloom\altaffilmark{15,16},
Peter~E.~Nugent\altaffilmark{15,16},
Mansi~M.~Kasliwal\altaffilmark{17,$\star$,$\dagger$},
Daniel~A.~Perley\altaffilmark{11,$\star$},
Robert~M.~Quimby\altaffilmark{18},
Adam~A.~Miller\altaffilmark{11,19,$\star$},
Christopher~A.~Theissen\altaffilmark{20}
and
Russ~R.~Laher\altaffilmark{21}
}

\affil{\altaffilmark{1}Department of Particle Physics and Astrophysics, The Weizmann Institute of Science, Rehovot 76100, Israel; \myemail}
\affil{\altaffilmark{2}Las Cumbres Observatory Global Telescope, 6740 Cortona Dr, Suite 102, Goleta, CA 93111, USA}
\affil{\altaffilmark{3}Kavli Institute for Theoretical Physics, University of California, Santa Barbara, CA 93106, USA}
\affil{\altaffilmark{4}School of Physics and Astronomy, University of Southampton, Southampton, SO17 1BJ, UK}
\affil{\altaffilmark{5}Department of Physics (Astrophysics), University of Oxford, DWB, Keble Rd, Oxford OX1 3RH, UK}
\affil{\altaffilmark{6}Astrophysics Science Division, NASA Goddard Space Flight Center, Mail Code 661, Greenbelt, MD 20771, USA}
\affil{\altaffilmark{7}Joint Space Science Institute, University of Maryland, College Park, MD 20742, USA}
\affil{\altaffilmark{8}Infrared Processing and Analysis Center, California Institute of Technology, Pasadena, CA 91125, USA}
\affil{\altaffilmark{9}Key Laboratory for Research in Galaxies and Cosmology, The University of Sciences and Technology of China, Chinese Academy of Sciences, Hefei, Anhui 230026, China}
\affil{\altaffilmark{10}Department of Physics, University of California, Santa Barbara, CA 93106, USA}
\affil{\altaffilmark{11}Cahill Center for Astrophysics, California Institute of Technology, Pasadena, CA 91125, USA}
\affil{\altaffilmark{12}Dark Cosmology Centre, Niels Bohr Institute, University of Copenhagen, Juliane Maries Vej 30, 2100 K\o benhavn \O, Denmark}
\affil{\altaffilmark{13}Excellence Cluster Universe, Technische Universit\"{a}t M\"{un}chen,
Boltzmann St. 2, D-85748, Garching, Germany}
\affil{\altaffilmark{14}Max Planck Institute for Astrophysics, Karl Schwarzschild St. 1, D-85748 Garching, Germany}
\affil{\altaffilmark{15}Department of Astronomy, University of California, Berkeley, CA 94720, USA}
\affil{\altaffilmark{16}Lawrence Berkeley National Laboratory, 1 Cyclotron Rd, Berkeley, CA 94720, USA}
\affil{\altaffilmark{17}The Observatories, Carnegie Institution for Science, 813 Santa Barbara St, Pasadena, CA 91101, USA}
\affil{\altaffilmark{18}Kavli IPMU (WPI), the University of Tokyo, 5-1-5 Kashiwanoha, Kashiwa-shi, Chiba 277-8583, Japan}
\affil{\altaffilmark{19}Jet Propulsion Laboratory, California Institute of Technology, Pasadena, CA, USA}
\affil{\altaffilmark{20}Astronomy Department, Boston University, 725 Commonwealth Ave, Boston, MA, USA}
\affil{\altaffilmark{21}Spitzer Science Center, California Institute of Technology, Pasadena, CA 91125, USA}

\altaffiltext{$\star$}{Hubble Fellow}
\altaffiltext{$\dagger$}{Carnegie-Princeton Fellow}




\begin{abstract} 

We present the results of a Palomar Transient Factory (PTF) archival search for blue transients which lie in the magnitude range between ``normal'' core-collapse and superluminous supernovae (i.e. with $-21\,{\leq}M_{R\,(peak)}\,{\leq}-19$). Of the six events found after excluding all interacting Type~IIn and Ia-CSM supernovae, three (PTF09ge, 09axc and 09djl) are coincident with the centers of their hosts, one (10iam) is offset from the center, and for two (10nuj and 11glr) a precise offset can not be determined. All the central events have similar rise times to the He-rich tidal disruption candidate PS1-10jh, and the event with the best-sampled light curve also has similar colors and power-law decay. Spectroscopically, PTF09ge is He-rich, while PTF09axc and 09djl display broad hydrogen features around peak magnitude. All three central events are in low star-formation hosts, two of which are E+A galaxies. Our spectrum of the host of PS1-10jh displays similar properties. PTF10iam, the one offset event, is different photometrically and spectroscopically from the central events and its host displays a higher star formation rate. Finding no obvious evidence for ongoing galactic nuclei activity or recent star formation, we conclude that the three central transients likely arise from the tidal disruption of a star by a super-massive black hole. We compare the spectra of these events to tidal disruption candidates from the literature and find that all of these objects can be unified on a continuous scale of spectral properties. The accumulated evidence of this expanded sample strongly supports a tidal disruption origin for this class of nuclear transients.

\end{abstract} 




\section{Introduction} 

\begin{figure}
\includegraphics[width=\columnwidth]{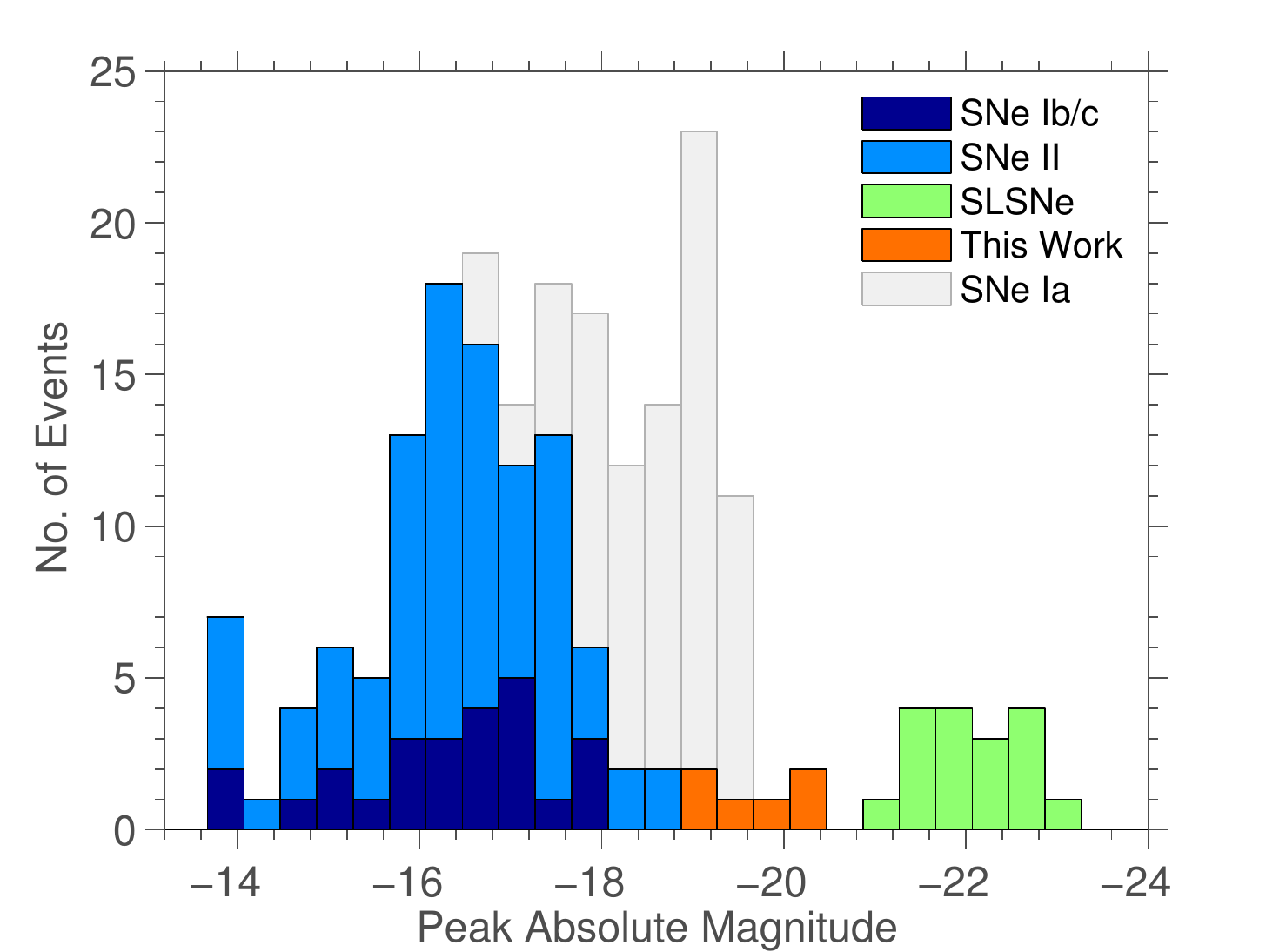}
\caption{\label{fig:peakhist}Peak magnitudes of Type Ia and core-collapse SNe from Li et al. (2011; see Filippenko 1997 for a review of SN types), SLSNe from Gal-Yam (2012), and the events presented in this work.}
\end{figure}

The peak luminosities of novae ($-10\,{\lesssim}\,M_R\,{\lesssim}-4$), supernovae (SNe; $-19\,{\lesssim}\,M_R\,{\lesssim}-14$) and superluminous SNe (SLSNe; $-21\,{\lesssim}\,M_R$; see Gal-Yam 2012 for a review) span a wide but discontinuous range. Discoveries of new types of transients (e.g. van Dyk et al. 2000; Valenti et al. 2009; Foley et al. 2009; Perets et al. 2010; Kasliwal et al. 2011) have been filling the luminosity gap between novae and SNe. The gap between SNe and SLSNe, however, is less explored (Fig. \ref{fig:peakhist}). Here we present the results of a search for such transients in the Palomar Transient Factory (PTF; Law et al. 2009; Rau et al. 2009) archive, focusing on events originally classified as core-collapse SNe.

\begin{figure}
\includegraphics[width=\columnwidth]{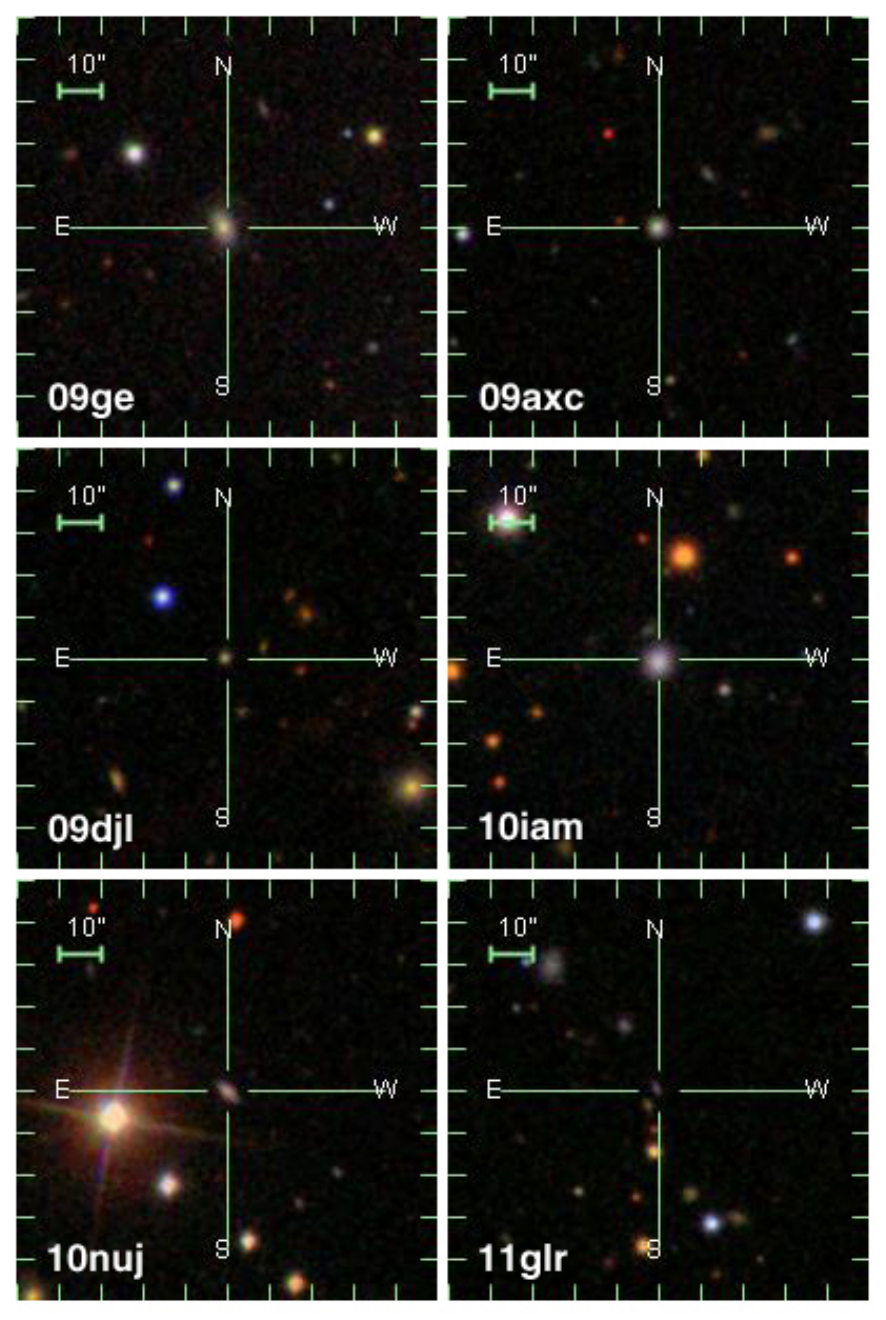}
\caption{\label{fig:sdss_hosts}SDSS DR10 images centered at the locations of the events listed in Table \ref{tab:events}.}
\end{figure}

PTF is an untargeted survey which discovered and classified over $500$ core-collapse SNe between the years 2009 and 2012 (iPTF\footnote{http://ptf.caltech.edu/iptf/} continues as the successor of this survey). Of those, we search for events originally classified as Type II SNe due to the presence of broad (a few thousand $\textrm{km\,s}^{-1}$) hydrogen lines or a blue continuum in their spectra, and displaying a peak absolute magnitude $-21\,{\leq}M_R\,{\leq}-19$. We exclude from this sample all Type IIn events (SNe showing narrow emission lines, e.g. Schlegel et al. 1990; Kiewe et al. 2012) and Ia-CSM events (e.g. Hamuy et al. 2003; Silverman et al. 2013), known to extend into this magnitude range (e.g. Stoll et al. 2011) likely due to emission powered by interaction with a dense circum-stellar medium. These events were typed using both visual and automatic classifications carried out with the SN spectral fitting codes Superfit (Howell et al. 2005) and SNID (Blondin \& Tonry 2007). We are left with the six events whose spectra do not match those of known types of interacting SNe. These events are presented in Table \ref{tab:events} and Figure \ref{fig:sdss_hosts}.

\begin{table*}
{\caption{\label{tab:events}All PTF events originally classified as Type II SNe due to the presence of broad H or a blue continuum in their spectra, having a peak absolute magnitude between $-19$ and $-21$. Here we exclude interacting Type IIn and Ia-CSM events. Coordinates are based on P48 images astrometrically calibrated to SDSS (with typical errors of $0.1^{\prime\prime}$). Peak magnitudes refer to the brightest measured photometric point and its error. The offset of each target with respect to its host center was determined using relative image registration, and the normalized distance (ND) is noted (see Section \ref{sec:offsets} for details). We separate the sample into three classes: central events ($\textrm{ND}\leq1$), offset events ($\textrm{ND}\geq3$), and intermediate events ($1<\textrm{ND}<3$) for which we can not robustly conclude a coincidence or separation from their host center.}}
\begin{tabular}{lllllllllll}
\hline
\hline
{Name} & {Type} & {$\alpha$ (J2000)} & {$\delta$ (J2000)} & {Redshift} & {Discovery} & {Discovery} & {Peak $M_R$} & \multicolumn{2}{l}{Host Offset [mas]} & {ND} \tabularnewline
{} & {} & {} & {} & {} & {Date} & {$m_R$} & {} & {P48} & {SDSS} & {} \tabularnewline
\hline
{09ge} & {He} & {14:57:03.10} & {+49:36:40.8} & {0.064} & {2009 May 7} & {$19.22\pm0.05$} & {$-19.64\pm0.02$} & {$84\pm42$} & {$82\pm81$} & {1.0} \tabularnewline
{09axc} & {H} & {14:53:13.06} & {+22:14:32.2} & {0.1146} & {2009 June 20} & {$20.83\pm0.12$} & {$-19.53\pm0.04$} & {$119\pm60$} & {$78\pm84$} & {0.9} \tabularnewline
{09djl} & {H} & {16:33:55.94} & {+30:14:16.3} & {0.184} & {2009 July 24} & {$20.76\pm0.10$} & {$-20.20\pm0.10$} & {$175\pm155$} & {$84\pm109$} & {0.8} \tabularnewline
\hline
{10iam} & {H} & {15:45:30.85} & {+54:02:33.0} & {0.109} & {2010 May 22} & {$20.73\pm0.26$} & {$-20.13\pm0.02$} & {$929\pm67$} & {$934\pm81$} & {11.4} \tabularnewline
\hline
{10nuj} & {H} & {16:26:24.70} & {+54:42:21.6} & {0.132} & {2010 June 13} & {$21.29\pm0.17$} & {$-19.31\pm0.05$}  & {$304\pm68$} & {$217\pm82$} & {2.6} \tabularnewline
{11glr} & {H} & {16:54:06.13} & {+41:20:14.8} & {0.207} & {2011 May 28} & {$21.44\pm0.29$} & {$-19.85\pm0.08$}  & {$1199\pm1390$} & {$941\pm758$} & {1.2} \tabularnewline
\hline
\end{tabular}
\end{table*}

Several types of transients, not necessarily related to massive stars, could still fulfill all of our search criteria. One such example is the flare resulting from a tidal disruption event (TDE; Rees 1988). A TDE can occur when a star passes close enough to a super-massive black hole (SMBH) and is torn apart by gravitational tidal forces. For a star of mass $M_{*}$ and radius $R_{*}$, and a SMBH with mass $M_{BH}$, this will occur at $R_T\sim\left(M_{BH}/M_{*}\right)^{1/3}R_{*}$. Part of the disrupted star is accreted onto the SMBH and part is unbound. If the mass of the SMBH satisfies\footnote{using a Newtonian calculation for a non-spinning black hole (see e.g. Kesden 2012 for a relativistic derivation yielding higher limits for rotating black holes)}
\begin{equation}
M_{BH}\lesssim10^{8}M_{\odot}\cdot\left(\frac{R_{*}}{R_{\odot}}\right)^{3/2}\left(\frac{M_{*}}{M_{\odot}}\right)^{-1/2}
\end{equation} 
then the tidal disruption radius is greater than the SMBH event horizon and a flare of radiation is expected to be observed. The observational signature of this flare depends on the structure of the accreting debris disk and on the morphology of the ejected material (Ulmer 1999; Bogdanovic et al. 2004; Strubbe \& Quataert 2009; Guillochon et al. 2014).

The expected rate of TDEs lies in the range of $10^{-5}$ to $10^{-4}$ events per galaxy per year (e.g. Donley et al. 2002; Wang \& Merritt 2004; Kesden 2012; but see also Alexander 2012 for a discussion on sources of uncertainty for this rate). A few TDE candidates were found in X-ray data from \emph{ROSAT} (Komossa \& Bade 1999; Donley et al. 2002; Halpern et al. 2004), \emph{Chandra} (Komossa et al. 2004) and \emph{XMM-Newton} (Esquej et al. 2007), in $\gamma$-ray and X-ray data from \emph{Swift} (Bloom et al. 2011; Burrows et al. 2011; Levan et al. 2011; Zauderer et al. 2011; Cenko et al. 2012), in UV data from \emph{GALEX} (Gezari et al. 2006; Gezari et al. 2009) and in optical data from the Sloan Digital Sky Survey (SDSS\footnote{http://www.sdss3.org/}; van Velzen et al. 2011). The high energy emission is associated with a jet pointing in our line of sight, while the soft X-ray/UV-optical flare is usually identified with thermal emission coming from an accretion flow. 

One expected observational signature of such accretion-powered emission is a light curve decline rate of $t^{-5/3}$ which follows from a $t^{-5/3}$ expected decay in the accretion rate (Rees 1988; Evans \& Kochanek 1989; Phinney 1989)\footnote{This light curve decline rate is similar to that of Type IIL SNe, but those events typically peak at lower luminosities (Arcavi et al. 2012) compared to what is expected for a TDE.}. The $t^{-5/3}$ mass accretion rate assumes a uniform energy distribution for the post-disrupted material. Lodato et al. (2009) show that this assumption fails for certain stellar density profiles, and that TDE light curves may deviate from $t^{-5/3}$ at early times. An expanding disk may alter this rate also at late times (Shen \& Matzner 2014), and it may be entirely different for partial disruptions (Guillochon \& Ramirez-Ruiz, 2013). Finally, Strubbe \& Quataert (2009) warn that even if the mass accretion rate does decay as $t^{-5/3}$, the flux in a given band will not necessarily follow the same rule.

Gezari et al. (2012; hereafter G12) presented the joint discovery of PS1-10jh in the optical and UV by Pan-STARRS1 (PS1) and the GALEX Time Domain Survey (TDS), respectively, and identified it as a likely TDE. The event is a luminous (bolometric $L_{\rm peak}\sim10^{44}$ erg~s$^{-1}$) and blue ($T_{\rm blackbody}\sim30000$ K) transient declining on a time scale of a few months. It is the first TDE candidate to have its rise to peak well sampled, and the light curve is roughly consistent with the models of Lodato et al. (2009). 

Chornock et al. (2014) report on the PS1 and GALEX TDS discovery of PS1-11af, which is similar to PS1-10jh but with a lower effective temperature. They find good fits to models with an accreted mass as small as $0.002M_{\odot}$, leading them to suggest that PS1-11af was a partial TDE (i.e. the star was not fully disrupted). Their spectra show no clear features except for a broad UV absorption component (possibly associated with Mg II), but their wavelength coverage can not be used to rule out any H${\alpha}$ emission. PS1-10jh, in contrast, displayed broad emission lines at rest-frame wavelengths of $3203\,{\textrm{\AA}}$ and $4686\,{\textrm{\AA}}$, interpreted as He II, with no obvious signs of hydrogen. 

G12 explain the lack of observed hydrogen in PS1-10jh with stellar winds or stripping of the star during previous passages near the SMBH (Bogdanovic et al. 2013 discuss the plausibility of the latter scenario). Guillochon et al. (2014) claim that hydrogen would not be visible at early times even if it were present in the disrupted star. In their model, the optical emission is dominated by the bound material in a regime where the hydrogen is fully ionized, and therefore not observable until long after peak.

Wang et al. (2011) searched for TDE candidates in SDSS by looking for narrow high-ionization coronal lines. The spectrum of one of their objects, SDSS J074820.66+471214.6 (hereafter SDSS J0748), showed also broad emission features around He II $4686\,{\textrm{\AA}}$ {\it and} around H${\alpha}$. Yang et al. (2013) re-observed SDSS J0748 several years after the initial SDSS spectrum and found that the broad features had disappeared, confirming their transient nature. However, with no light curve for this event, it is hard to compare it to PS1-10jh directly.

PS1-10jh was also detected by PTF as PTF10onn and marked as a possible TDE by the autonomous software framework Oarical (Bloom et al. 2012). With a peak magnitude of $M_r=-19.5$, it would have come up in the archival search discussed here, but a spectrum was never obtained as part of PTF followup. We do, however, find very similar events. One (PTF09ge; Kasliwal et al. 2009) displays a nearly identical He II emission feature as SDSS J0748 but with no hydrogen and with a light curve very similar to PS1-10jh. Two more of the events found in our search (PTF09axc and PTF09djl) show broad hydrogen features starting from the earliest spectra (taken a few days after peak magnitude).

These two H-rich events happen to be located in the centers of rare E+A galaxies (Dressler \& Gunn 1983). Such galaxies show no emission lines that are indicative of on-going or recent star formation. They are thus not likely to host core-collapse SNe which originate in massive, short-lived stars. Our spectrum of the host galaxy of PS1-10jh shows similar features. E+A galaxies (sometimes referred to as K+A) are so-called because the Balmer absorption features in their spectra (characteristic of A-stars) appear superimposed on an old K-star or (E)arly type galaxy population. The A-stars would have been formed in an episode of star formation which ceased abruptly $\sim1$ Gyr ago (Dressler \& Gunn 1983), possibly following a merger. 

Recently, Prieto et al. (2014) and Holoien et al. (2014) reported the discovery of ASASSN-14ae as a likely TDE. We find that it displays spectral features similar to those seen in SDSS J0748, with broad emission in both He II and H.

We note that another type of transient expected to occur exclusively in galaxy centers was recently suggested by Bablerg et al. (2013). They propose that hypervelocity stellar collisions could result in bright SN-like transients. However, their prediction for the peak magnitudes of such flares are at the lower end of the known SN luminosity scale, much fainter than the events discussed here.

We present photometric and spectroscopic data of all six events found in our archival search in \S 2, and analyze these observations in \S 3. In \S 4 we briefly discuss the events which could not be robustly associated with the center of their host. We then focus on the three central events in \S 5, comparing them to the TDE candidates mentioned above. We conclude in \S 6.

\section{Observations}

All the events from our archival search were discovered by the Palomar 48-inch Oschin Schmidt Telescope (P48), as part of the PTF survey, using the Mould $R$-band filter. The discovery details are presented in Table \ref{tab:events}.

\subsection{Photometry}

\begin{figure}
\includegraphics[trim=0 0 0 50, clip,width=\columnwidth]{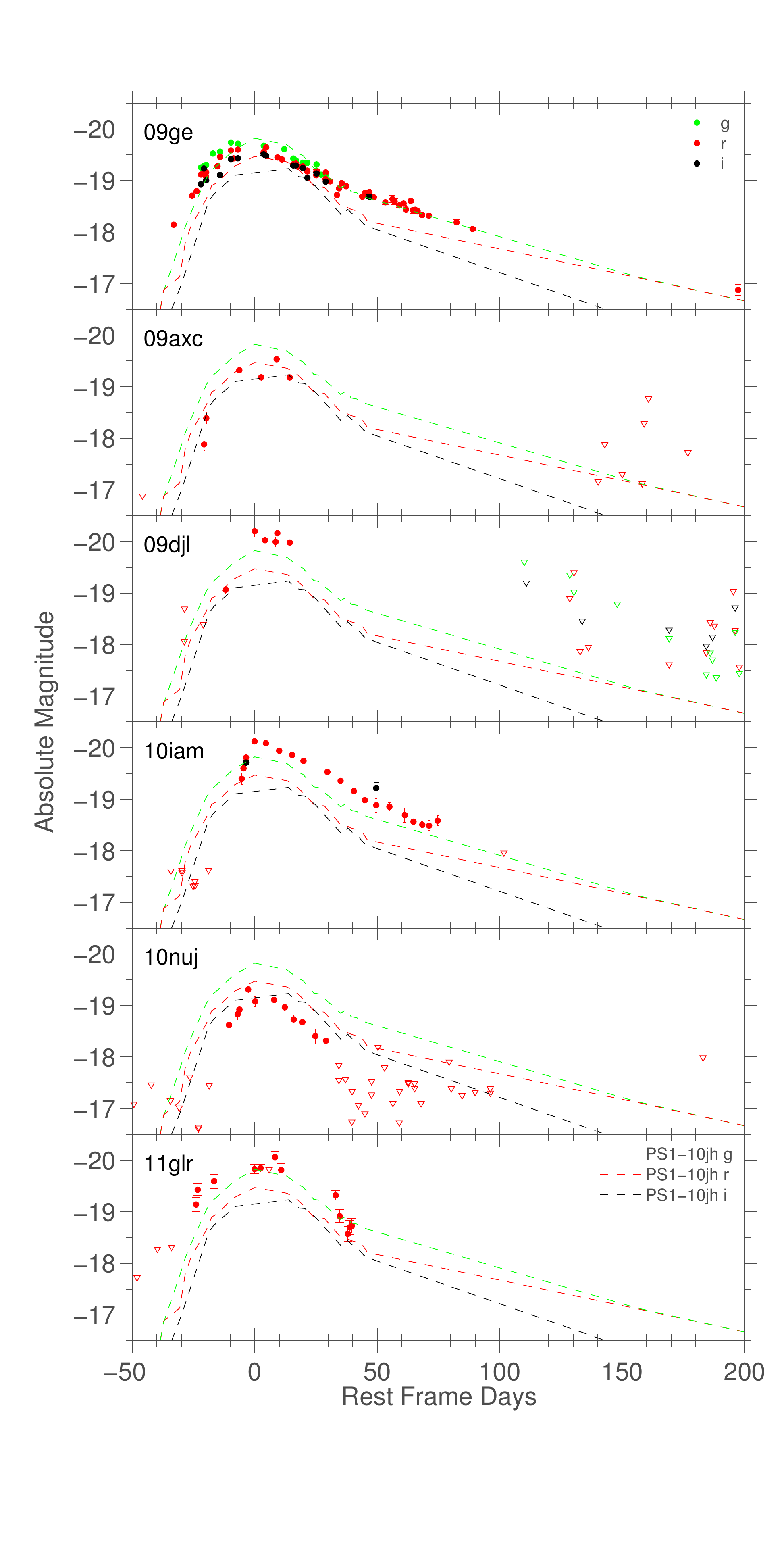}\vspace{-10 mm}
\caption{\label{fig:all_phot}Light curves of PTF09ge, PTF09axc, PTF09djl, PTF10iam, PTF10nuj and PTF11glr from P48 and P60. Most display light curve shapes consistent with those of PS1-10jh (G12; dashed lines). PTF10iam shows a uniquely fast rise to peak, and PTF10nuj shows a sudden luminosity drop $\sim30$ days after peak.}
\end{figure}

We obtained photometric observations in the $R$ and $g$-bands using P48, and in $g$, $r$ and $i$-bands with the Palomar 60-inch telescope (P60; Cenko et al. 2006). Initial processing of the P48 images was conducted by the Infrared Processing and Analysis Center (IPAC; Laher et al. 2014). Photometry was extracted using a custom PSF fitting routine (e.g. Sullivan et al. 2006), which measures the transient flux after image subtraction (using template images taken before the outburst or long after it faded). We calibrate our light curves to the SDSS system using SDSS observations (Ahn et al. 2012) of the same field when possible, and to USNO-B (Monet et al. 2003) reference magnitudes otherwise. We average magnitudes obtained from the same filter and instrument taken on the same night. We correct for Galactic extinction using the Schlafly \& Finkbeiner (2011) maps, extracted via the NASA Extragalctic Database (NED\footnote{http://ned.ipac.caltech.edu/}). Distance moduli are calculated from spectroscopic redshifts, measured using narrow host features. A cosmological model with $H_0 = 70\,\textrm{km\,s}^{-1}\,\textrm{Mpc}^{-1}$, $\Omega_{m}=0.3$ and $\Omega_{\Lambda}=0.7$ is assumed throughout. Our photometry is presented in the AB system in Table \ref{tab:phot} and Figure \ref{fig:all_phot}.

Due to gaps in the photometry, we note that all phases stated hereafter (regarding the PTF sample) should be considered with few-day uncertainties.

\begin{table}
{\caption{\label{tab:phot}Photometric observations (upper limits mark $3\sigma$ non-detections). This table is published in its entirety in the electronic version. A portion is shown here for guidance regarding its form and content.}}
\begin{tabular}{llllll}
\hline
\hline
{Object} & {Telescope} & {Filter} & {JD} & {Mag} & {Error}\tabularnewline
\hline
{09ge} & {P48} & {g} & {$2454910.954$} & {$>22.001$} & {}\tabularnewline
{09ge} & {P48} & {g} & {$2454917.759$} & {$>21.349$} & {}\tabularnewline
{09ge} & {P48} & {g} & {$2454918.904$} & {$>21.597$} & {}\tabularnewline
{09ge} & {P48} & {g} & {$2454921.74$} & {$>21.384$} & {}\tabularnewline
{09ge} & {P48} & {g} & {$2454921.817$} & {$>21.532$} & {}\tabularnewline
{09ge} & {P48} & {g} & {$2454975.742$} & {$17.833$} & {$0.008$}\tabularnewline
{09ge} & {P48} & {g} & {$2454975.803$} & {$17.852$} & {$0.011$}\tabularnewline
\hline
\end{tabular}
\end{table}

\subsection{AO Imaging}

We imaged the host galaxy of PTF09djl on 2013~July~5 (several years after the transient faded) using the Laser Guide Star Adaptive Optics system (Wizinowich 2006) and the NIRC2 camera on the Keck II 10-meter telescope in the $K_p$ band with the $40^{\prime\prime}$ square field of view ``wide'' camera. 

The image frames were dark-subtracted and flat-fielded in the standard manner using custom Python scripts. The effects of fringing were removed by subtracting a sky fringe scaled to the sky brightness of each frame. Since 2007, a variable glow has been present at the lower right corner of the ``wide'' camera that cannot be adequately removed by calibration. A triangular section of the frames was masked before registering and coadding the images. The resulting image is shown in Figure \ref{fig:ao_reg}.

\subsection{Spectroscopy}

\begin{figure}
\includegraphics[trim=0 0 0 50, clip,width=\columnwidth]{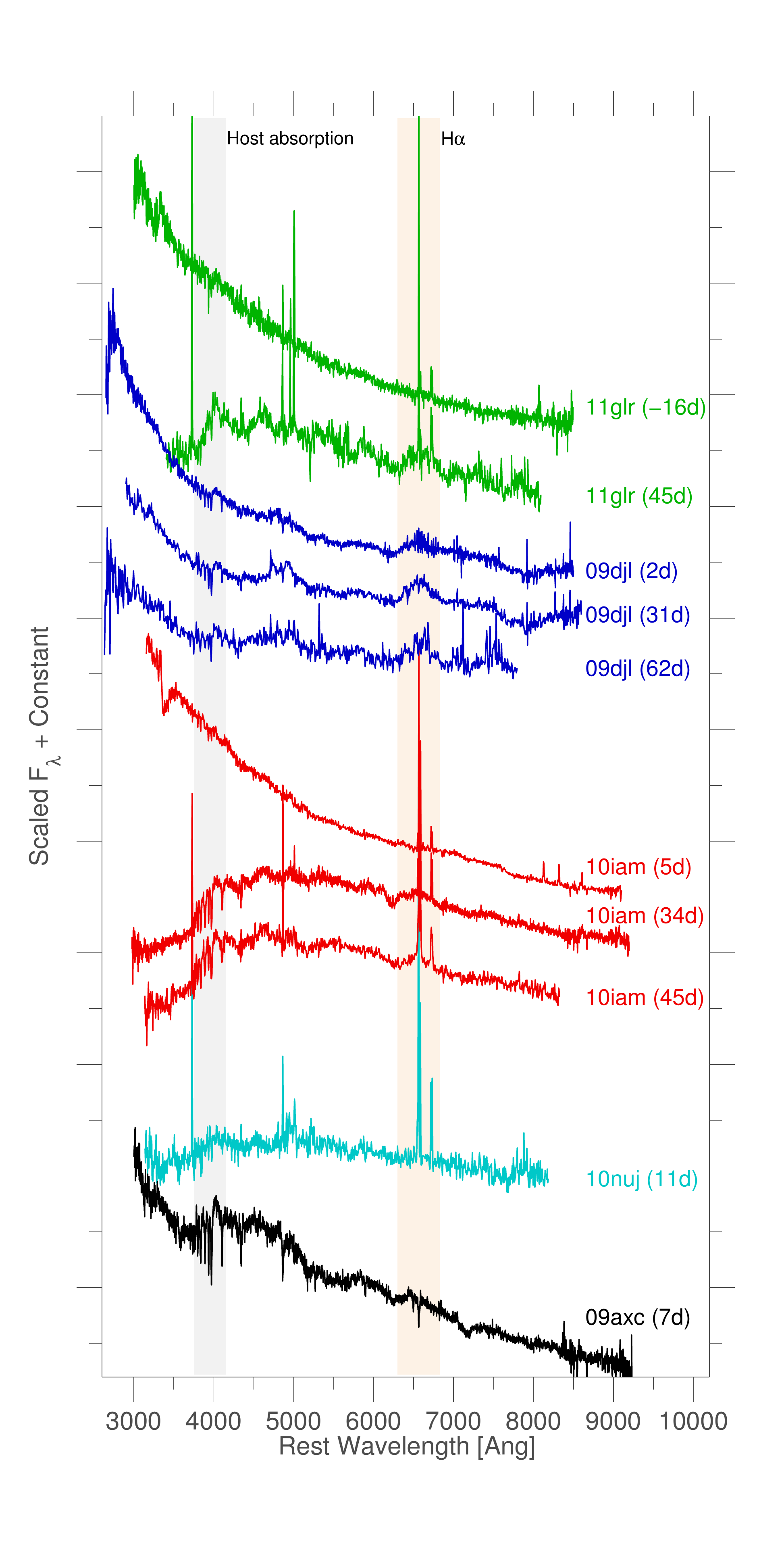}\vspace{-6 mm}
\caption{\label{fig:all_specs}Spectra of PTF09axc, PTF09djl, PTF10iam, PTF10nuj and PTF11glr. Phases are shown relative to peak. Host absorption features and the broad H$\alpha$ region are marked with thick bands.}
\end{figure}

Spectra of the events were obtained with the Double Beam Spectrograph (DBSP; Oke \& Gunn 1982) mounted on the Palomar 200-inch telescope (P200) and the Low Resolution Imaging Spectrometer (LRIS; Oke et al. 1995) mounted on the Keck I 10-meter telescope (Table \ref{tab:spec}). The data were reduced using standard IRAF\footnote{IRAF, the Image Reduction and Analysis Facility, is a general purpose software system for the reduction and analysis of astronomical data. IRAF is written and supported by the National Optical Astronomy Observatories (NOAO) in Tucson, Arizona} and IDL routines. Host galaxy spectra were obtained in 2013, after all transient emission had faded (except for PTF10iam, for which a host spectrum was obtained by the SDSS in 2002 and downloaded via DR10; Ahn et al. 2013). Spectra of the transients are presented in Figures \ref{fig:all_specs} and \ref{fig:09ge_spec}, and host galaxy spectra are shown in Figure \ref{fig:host_specs}. Digital versions of our spectra are available online through the Weizmann Interactive Supernova data REPository (WISeREP\footnote{http://www.weizmann.ac.il/astrophysics/wiserep}; Yaron \& Gal-Yam 2012).

\begin{table}
{\caption{\label{tab:spec}Spectroscopic observations. The phase is denoted in days relative to $R$-band maximum.}}
\begin{tabular}{lllll}
\hline
\hline
{Object} & {UT Date} & {Phase} & {Telescope} & {Instrument}\tabularnewline
\hline
{09ge} & {2009 May 20} & {-24} & {P200} & {DBSP}\tabularnewline
{09axc} & {2009 July 22} & {7} & {Keck I} & {LRIS}\tabularnewline
{09djl} & {2009 Aug 25} & {2} & {Keck I} & {LRIS}\tabularnewline
{09djl} & {2009 Sep 23} & {31} & {Keck I} & {LRIS}\tabularnewline
{09djl} & {2009 Oct 24} & {62} & {Keck I} & {LRIS}\tabularnewline
{10iam} & {2010 June 8} & {5} & {Keck I} & {LRIS}\tabularnewline
{10iam} & {2010 July 7} & {34} & {Keck I} & {LRIS}\tabularnewline
{10iam} & {2010 July 18} & {45} & {P200} & {DBSP}\tabularnewline
{10nuj} & {2010 July 14} & {11} & {P200} & {DBSP}\tabularnewline
{11glr} & {2011 June 29} & {-16} & {Keck I} & {LRIS}\tabularnewline
{11glr} & {2011 Aug 28} & {45} & {Keck I} & {LRIS}\tabularnewline
\hline
{09ge Host} & {2013 Sep 9} & {} & {Keck I} & {LRIS}\tabularnewline
{09axc Host} & {2013 May 9} & {} & {Keck I} & {LRIS}\tabularnewline
{09djl Host} & {2013 May 9} & {} & {Keck I} & {LRIS}\tabularnewline
{10nuj Host} & {2013 May 9} & {} & {Keck I} & {LRIS}\tabularnewline
{11glr Host} & {2013 Oct 4} & {} & {Keck I} & {LRIS}\tabularnewline
\hline
\end{tabular}
\end{table}

\subsection{Radio}

We observed the field of PTF09axc with the Jansky Very Large Array (VLA) on June 28, 2014.  The observations were performed at both $3.5$\,GHz (S-band) and $6.1$\,GHz (C-band) in the D configuration. We used J1513+2388 as a phase calibrator and 3C286 as a flux calibrator. The data were analyzed using standard calibration scripts within the CASA software (McMullin et al. 2007).

\section{The Sample}

\subsection{Offsets from Host Centers}
\label{sec:offsets}

After registering the images of each event, we measure its position from a host-subtracted image taken near peak magnitude, and compare it to the host centroid measured in a co-added reference image, using the IRAF task \textsc{imexamine}. We estimate the error in the measured offset of the transient using the scatter in centroid determinations (when varying the fit type and radii parameters of \textsc{imexamine} between 2 and 7 pixels) and the registration error (typically $\lesssim0.03$ pixels, corresponding to $\lesssim0.03^{\prime\prime}$ in the P48 images). 

We also measure the host galaxy centroids in the SDSS $r$-band images and register them to the P48 images. Here the registration errors are closer to $\sim0.06^{\prime\prime}$, but the finer pixel-scale of the SDSS images contributes to a more accurate host centroid determination for the brighter hosts. We repeat these measurements for the P48 photometric data of PS1-10jh, for comparison.

\begin{figure}
\includegraphics[width=\columnwidth]{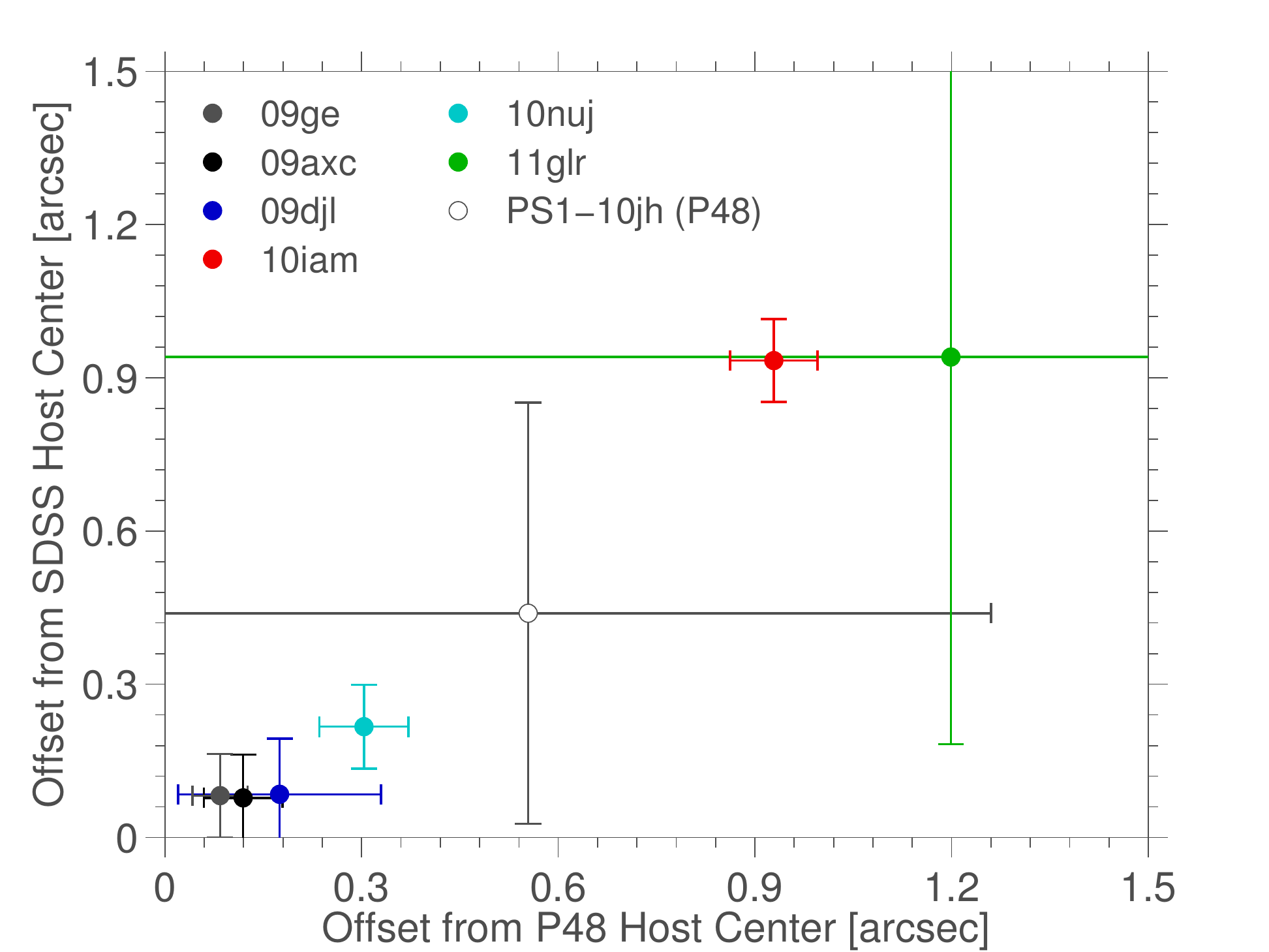}
\includegraphics[width=\columnwidth]{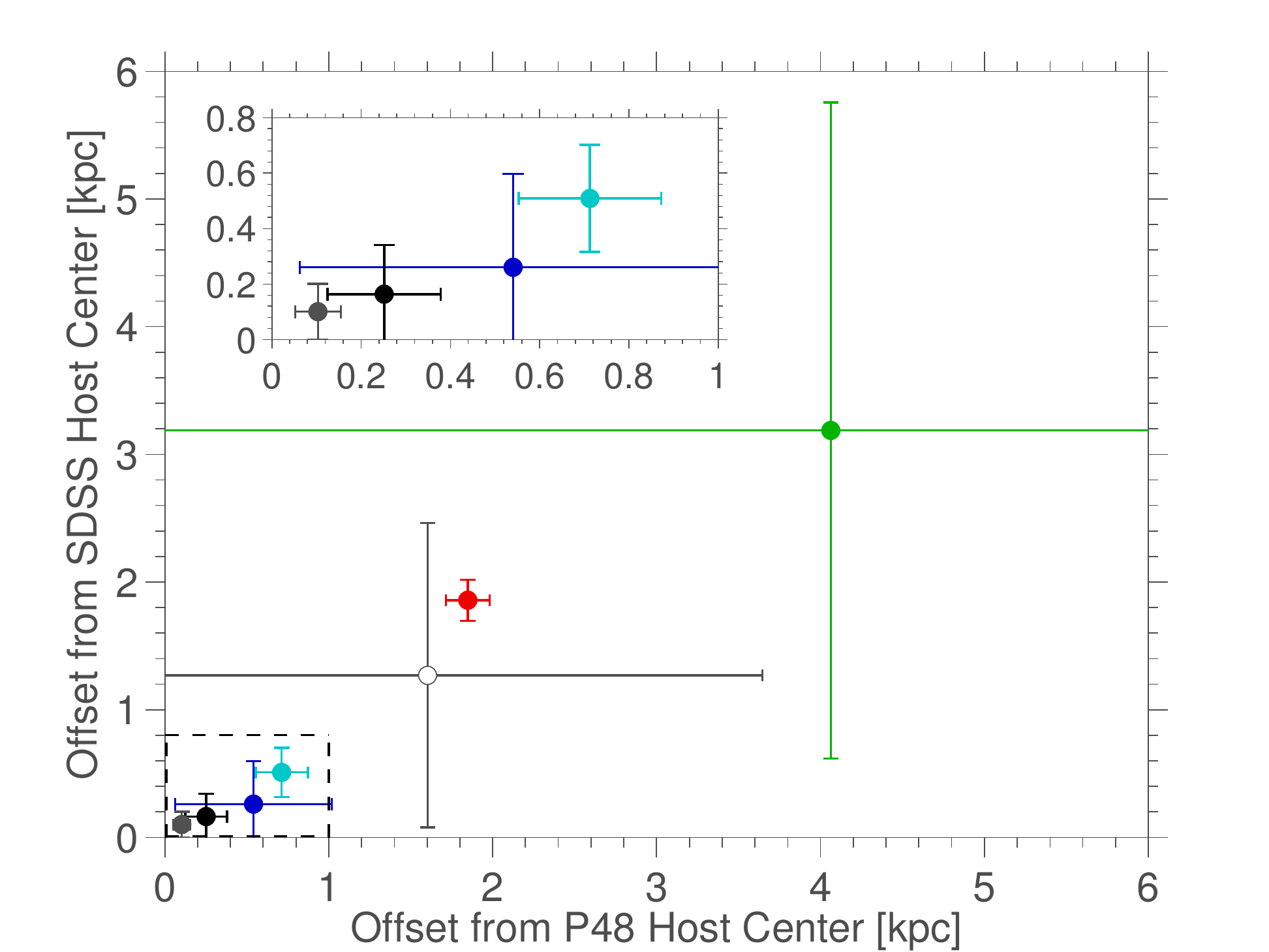}
\caption{\label{fig:offsets}Offsets in arcseconds (top) and kpc (bottom) between the locations of our events and the centers of their hosts, determined both from P48 and from SDSS images of the hosts registered to P48 images of the transients. Error bars denote $1\sigma$ uncertainties. The region enclosed in the dashed box in the bottom panel is shown in greater detail in the inset. Three events (PTF09ge, PTF09axc and PTF09djl) are found to be coincident with the centers of their hosts with relatively small errors, one event (PTF10iam) is found to be offset from its host center, and for the remaining two events (PTF10nuj and PTF11glr) it is not possible to determine either an offset nor a robust coincidence with the center. The position of PS1-10jh (from its P48 detection as PTF10onn) is shown for comparison.}
\end{figure}

The offsets of our events (as well as that of PS1-10jh), using the comparison to both P48 host centers and SDSS host centers, are presented in Figure \ref{fig:offsets}.

We normalize each apparent offset to its galaxy \emph{expRad\_r} parameter (an estimate of the half-light radius of the galaxy) from SDSS DR10. We then divide each normalized offset by its error, and call this parameter ND (for ``Normalized Distance''). ND measures how many sigma each event is from its host center, in terms of its half-light normalized distance. The results are shown in Table \ref{tab:events}.

We find that one event (PTF10iam), with $\textrm{ND}=11.4$, is obviously offset from the center of its host. Three events (PTF09ge, PTF09axc and PTF09djl) are coincident with the centers of their hosts within the errors ($\textrm{ND}\leq1$). For the two remaining events (PTF10nuj and PTF11glr), with $1<\textrm{ND}<3$, we do not conclude whether they are coincident with their host centers or not.

\begin{figure}
\includegraphics[width=\columnwidth]{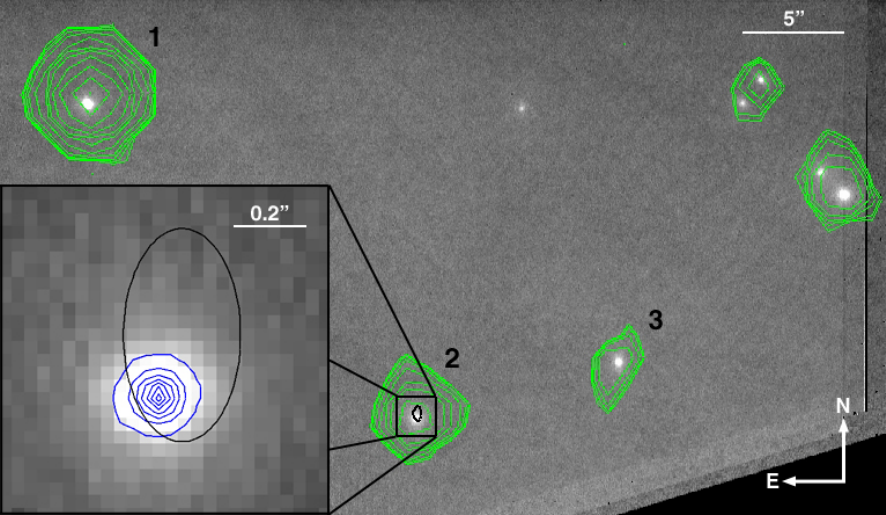}
\caption{\label{fig:ao_reg}NIRC2 AO image of the position of PTF09djl taken on 2013 July 5 (after the transient was no longer visible). The contours from the P48-resampled SDSS image are shown in green after WCS-registration and shifting to match the centroids of the numbered sources. The P48 position of PTF09djl, taking into account the centroid and shifting errors ($1\sigma$), is shown in the black ellipse. \textsc{Inset}: Enlarged area around the host of PTF09djl, with the AO image contours shown in blue and the position of PTF09djl marked in black.\\}
\end{figure}

For PTF09djl, we use the AO imaging of its host to further constrain its position relative to the host center. We find a WCS solution for the PTF09djl AO image using Aladin (Bonnarel et al. 2000) by comparing to 2MASS stars in the field, and register this image to the P48-resampled SDSS image based on its WCS solution. We then adjust the registration using a mean offset of three objects in the field (one of which is the host galaxy of PTF09djl itself; see Figure \ref{fig:ao_reg}). We superimpose the SN position from P48 onto this registered AO image, taking into account the centroid errors and offsetting errors. The results are shown in the inset of Figure \ref{fig:ao_reg}. We conclude that the position of PTF09djl remains consistent with the center of its host also under this analysis.

\subsection{Blackbody Fits}

\begin{figure}
\includegraphics[width=\columnwidth]{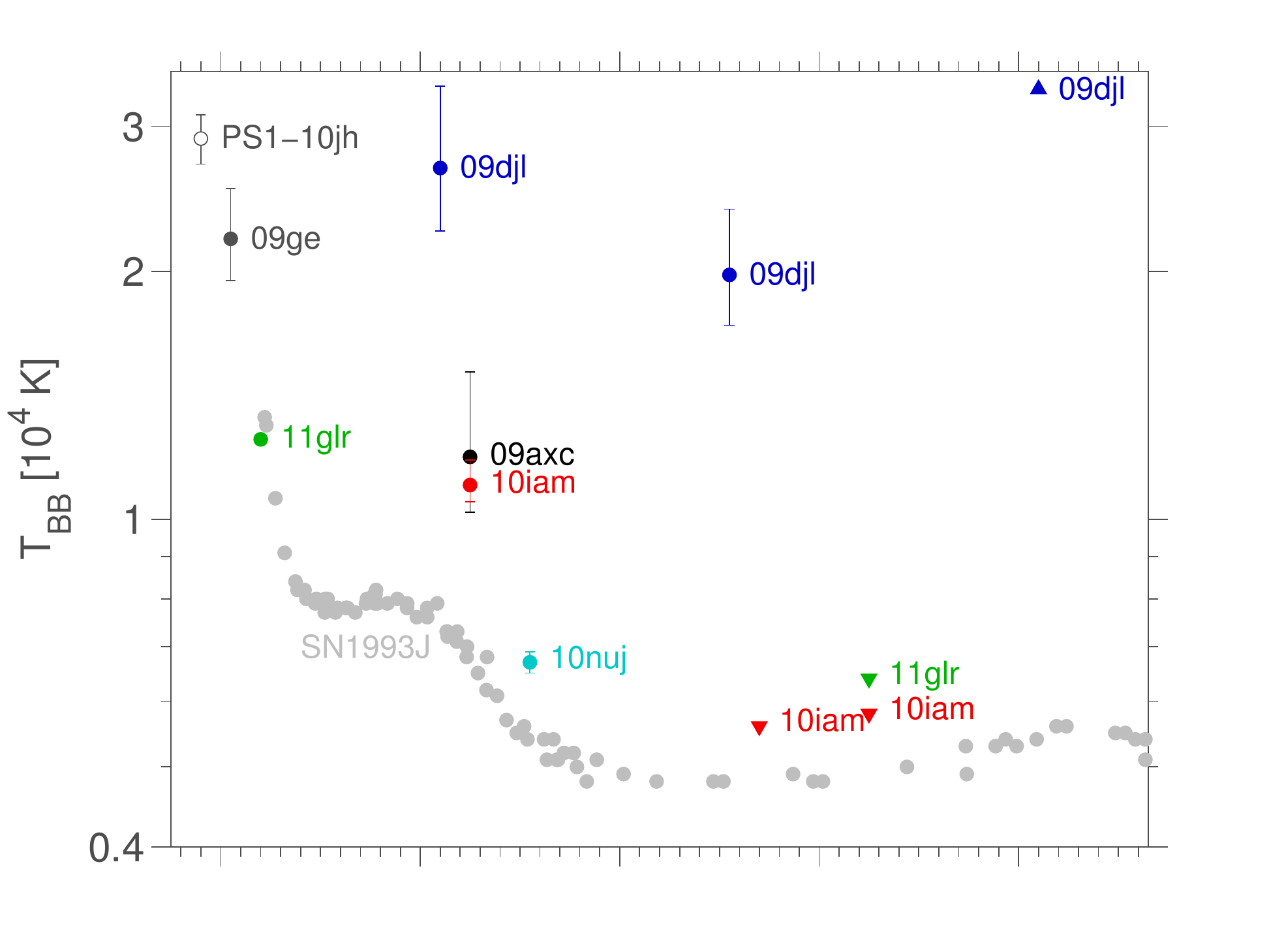}\vspace{-6 mm}
\includegraphics[width=\columnwidth]{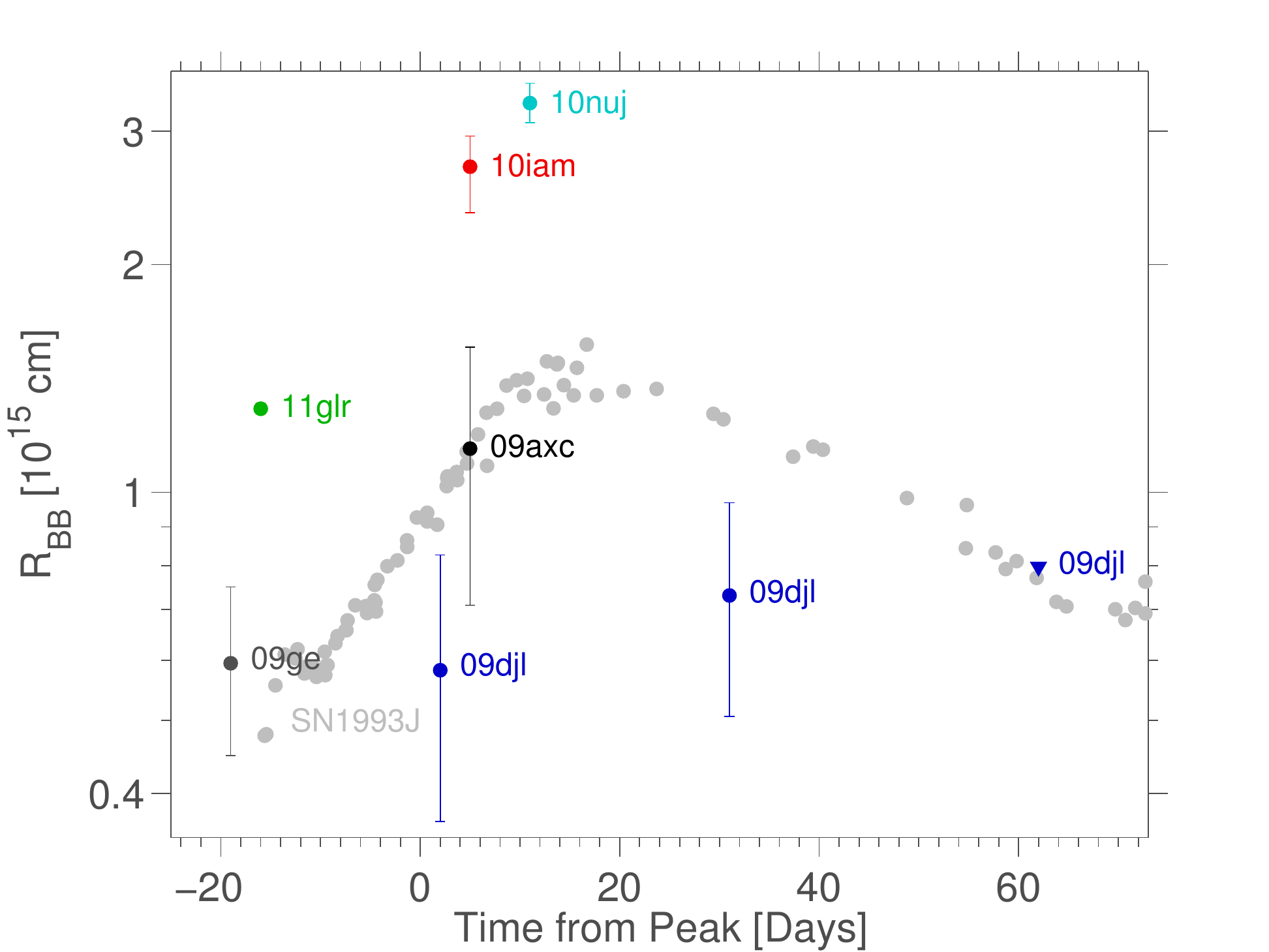}
\caption{\label{fig:bb}Top: Best fit black-body temperatures (when a good fit was possible) obtained from the optical host-subtracted spectra (triangles denote upper or lower limits). No host-extinction correction is performed. The black-body temperatures of PS1-10jh (G12) and SN1993J (Richmond et al. 1994) assuming low extinction are shown for comparison. Bottom: Blackbody radii estimated from the fits.}
\end{figure}

We fit a combination of a scaled host galaxy spectrum and a blackbody function to the spectra of each event. The best fit effective temperatures and radii (for blackbody fits yielding $T>6000$K, so as not to be dominated by lines over the continuum) are displayed in Figure \ref{fig:bb}. We are not able to correct for unknown host extinction, and therefore consider the measured temperatures to be lower limits.

Two of the central events, PTF09ge and PTF09djl, display high effective temperatures (consistent with PS1-10jh) and smaller radii compared the rest of the sample. The non-central events are cooler initially and continue to cool with time. The third central event, PTF09axc, displays an intermediate temperature and radius.

\subsection{\label{hosts}Host Galaxies}

\begin{figure*}
\includegraphics[trim=50 0 50 0, clip, width=\textwidth]{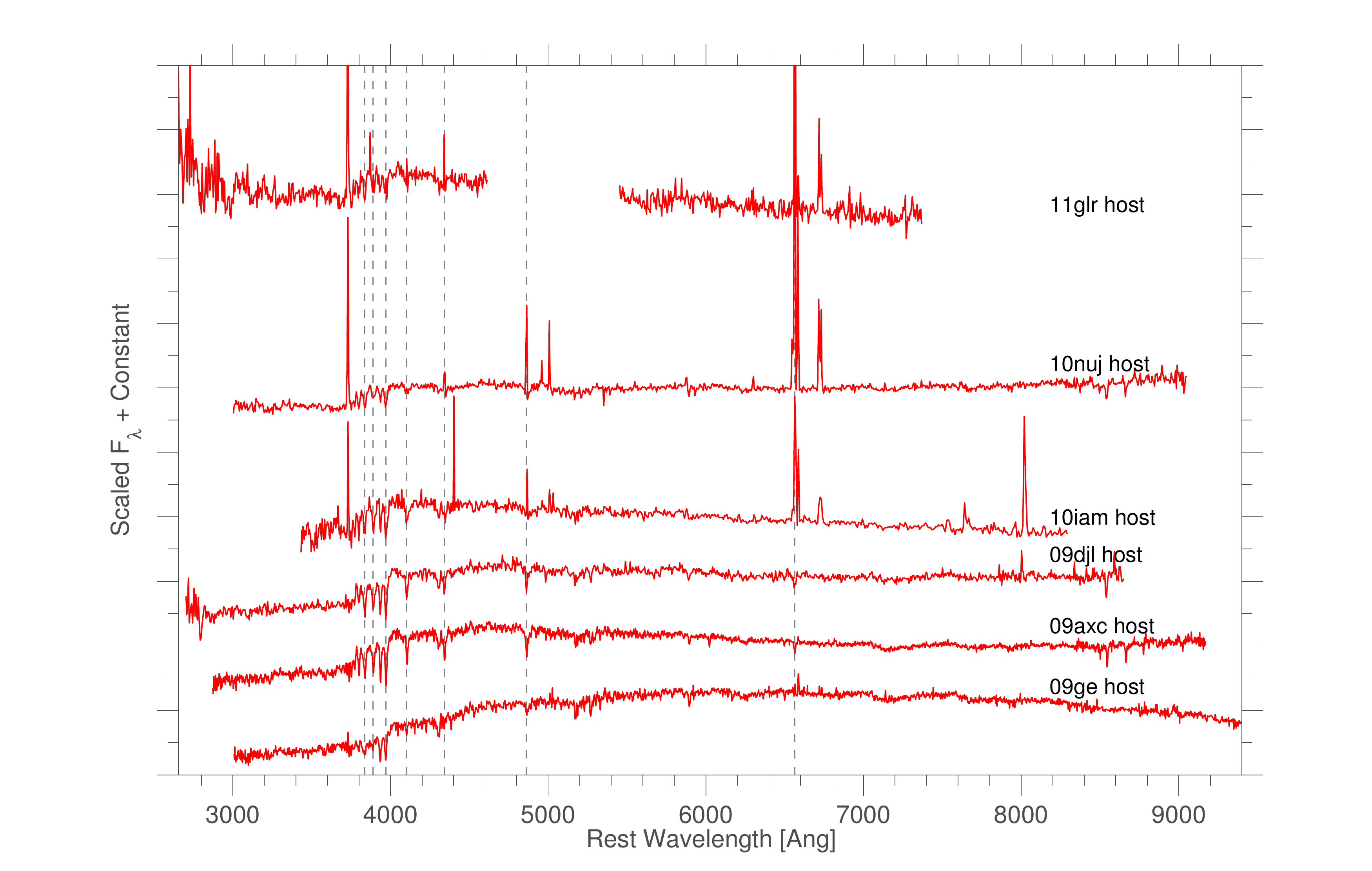}
\caption{\label{fig:host_specs}Host galaxy spectra for our archival sample (that of PTF10iam is from SDSS DR10). The hydrogen Balmer series, evident in absorption in E+A galaxies, is marked (dashed lines). The hosts of PTF09ge, 09axc and 09djl (the same events coincident with the centers of their hosts) do not show any strong emission lines, indicating no (or very low) on-going star formation, and display spectra similar to those of E+A galaxies.}
\end{figure*}

We obtain $ugriz$ magnitudes of all host galaxies from SDSS DR10 and analyze them using \textsc{z-peg} (Le Borgne \& Rocca-Volmerange 2002), which is based on the spectral synthesis code PEGASE.2 (Fioc \& Rocca-Volmerange 1997). \textsc{z-peg} fits the observed galaxy magnitudes with SED templates (of SB, Im, Sd, Sc, Sbc, Sb, Sa, S0 and E spectral types), to obtain the stellar mass and star-formation rate. We assume a Salpeter (1955) initial mass function and fit also a foreground dust screen varying in color excess from $E(B-V)=0$ to $0.2$ magnitudes. The results are presented in Table \ref{tab:host_params}.

Spectra of the host galaxies are presented in Figure \ref{fig:host_specs}. All of them display Balmer absorption features. While the host galaxies of PTF10iam, PTF10nuj and PTF11glr show emission lines, those of PTF09axc and PTF09djl (which are two of the three events coincident with the centers of their hosts) display spectra similar to E+A galaxies (Dressler \& Gunn 1983), with no strong indications of on-going or recent star formation. The host of the third central event (PTF09ge), is similar to E+A galaxies but shows some H$\alpha$ emission. We do not see narrow coronal emission lines in any of our host spectra taken after the transient had faded (c.f. Wang et al. 2012; Yang et al. 2013).

We fit the host spectra to stellar templates provided by the \textsc{miles} empirical stellar library (S{\'a}nchez-Bl{\'a}zquez et al. 2006; Vazdekis et al. 2010), using \textsc{ppxf} (Cappellari \& Emsellem 2004) and \textsc{gandalf} (Sarzi et al. 2006). We measure the star-formation rate (SFR) from the H${\alpha}$ luminosity using the conversion of Kennicutt (1998). We calibrate the gas-phase metallicity in our sample using the PP04 `N2' method (Pettini \& Pagel 2004) following the procedure described in Kewley \& Ellison (2008). We further determine the mass-weighted stellar metallicity and age by fitting the full stellar continuum with \textsc{ppxf} and weight averaging all the stellar templates. The SFR determined for galaxies showing insignificant or no H$\alpha$ lines in their spectrum should be interpreted with caution here as it depends on the stellar absorption fitted by the model templates. For more details on these procedures see Pan et al. (2013). All best-fit parameters are presented in Table \ref{tab:host_params}.

\begin{table*}
\caption{\label{tab:host_params}Properties of the host galaxies of our archival sample, obtained using \textsc{z-peg} SED template fits to SDSS $ugriz$ photometry of the hosts and from the host spectral features using the same methods discussed by Pan et al. (2013). Values in parenthesis (when noted) describe the lower and upper limits deemed acceptable by \textsc{z-peg} or $1\sigma$ errors from the spectral analysis. Cases where no good fits were found are denoted by ``n/a''.}
\begin{tabular*}{\textwidth}{lllll @{\extracolsep{\fill}} lll}
\hline
\hline
{Host} & \multicolumn{3}{l}{Photometric Analysis} & \multicolumn{4}{l}{Spectroscopic Analysis} \tabularnewline
{} &  {M [$10^{10}M_{\odot}$]} & {SFR [$M_{\odot}\textrm{yr}^{-1}$]} & {sSFR [$10^{-10}\textrm{yr}^{-1}$]} & {SFR [$M_{\odot}\textrm{yr}^{-1}$]} & {$12+{\log}$(O/H)} & {[M/H]} & {Age [Gyr]} \tabularnewline
\hline
{09ge} & {$1.05$ ($1.03$, $1.35$)} & {n/a} & {n/a} & {$0.10$ ($0.05$)} & {$8.8732$ ($0.0638$)} & {$-0.196$} & {$7.035$}\tabularnewline
{09axc} & {$1.23$ ($1.16$, $1.28$)} & {$<16.11$} & {$<13.10$} & {$0.04$ ($0.02$)} & {n/a} & {$-0.356$} & {$4.469$}\tabularnewline
{09djl} & {$1.86$ ($1.07$, $3.73$)} & {$3.42$ ($1.22$, $4.19$)} & {$1.84$ ($0.34$, $3.02$)} & {n/a} & {n/a} & {$-0.218$} & {$4.461$}\tabularnewline
{10iam} & {$2.94$ ($2.83$, $4.39$)} & {$8.51$ ($6.53$, $9.40$)} & {$2.90$ ($1.59$, $3.02$)} & {$2.86$ ($1.41$)} & {$8.672$ ($0.233$)} & {$-1.114$} & {$5.717$}\tabularnewline
{10nuj} & {$1.22$ ($1.18$, $2.89$)} & {$3.86$ ($2.59$, $4.03$)} & {$3.16$ ($1.05$, $3.16$)} & {$8.11$ ($3.74$)} & {$8.698$ ($0.070$)} & {$-0.284$} & {$5.743$}\tabularnewline
{11glr} & {$0.30$ ($0.28$, $0.65$)} & {$1.09$ ($0.73$, $1.15$)} & {$3.58$ ($1.44$, $3.58$)} & {$1.22$ ($0.56$)} & {$8.305$ ($0.080$)} & {n/a} & {n/a}\tabularnewline
\hline
\end{tabular*}
\end{table*}

\section{The Non-Central Events}

PTF10iam is the only event in our sample for which a clear offset from the center of its host can be determined. It shows a faster rise to peak luminosity compared to the other events (Fig. \ref{fig:all_phot}), and it displays a broad absorption feature near rest wavelength $6200\,\textrm{\AA}$ (Fig. \ref{fig:all_specs}), not seen in the spectra of the central events. If interpreted as high velocity H$\alpha$, this feature may be an indication of interaction (Chugai et al. 2007) which does not manifest itself in narrow emission lines. In any case, it is clear that this off-center event is different photometrically and spectroscopically from the rest of the sample. We discuss PTF10iam in detail in a companion paper (Arcavi et al. {\it in prep}.)

For PTF10nuj and PTF11glr we are not able to measure or rule out an offset of their position relative to their host center. We leave the detailed analysis of the photometry and spectroscopy of these two events for a future publication.

\section{The Central Events as TDE Candidates}

We now focus on the three events coincident with the centers of their hosts: PTF09ge, PTF09axc and PTF09djl.

\subsection{\label{sec:agn}Could they be AGN?}

A transient associated with the center of a galaxy may be related to galactic nuclei activity. However, the spectra taken during the outbursts (and those taken several years later) do not show obvious emission lines typical of active galactic nuclei (AGNs).

\begin{figure}
\includegraphics[width=\columnwidth]{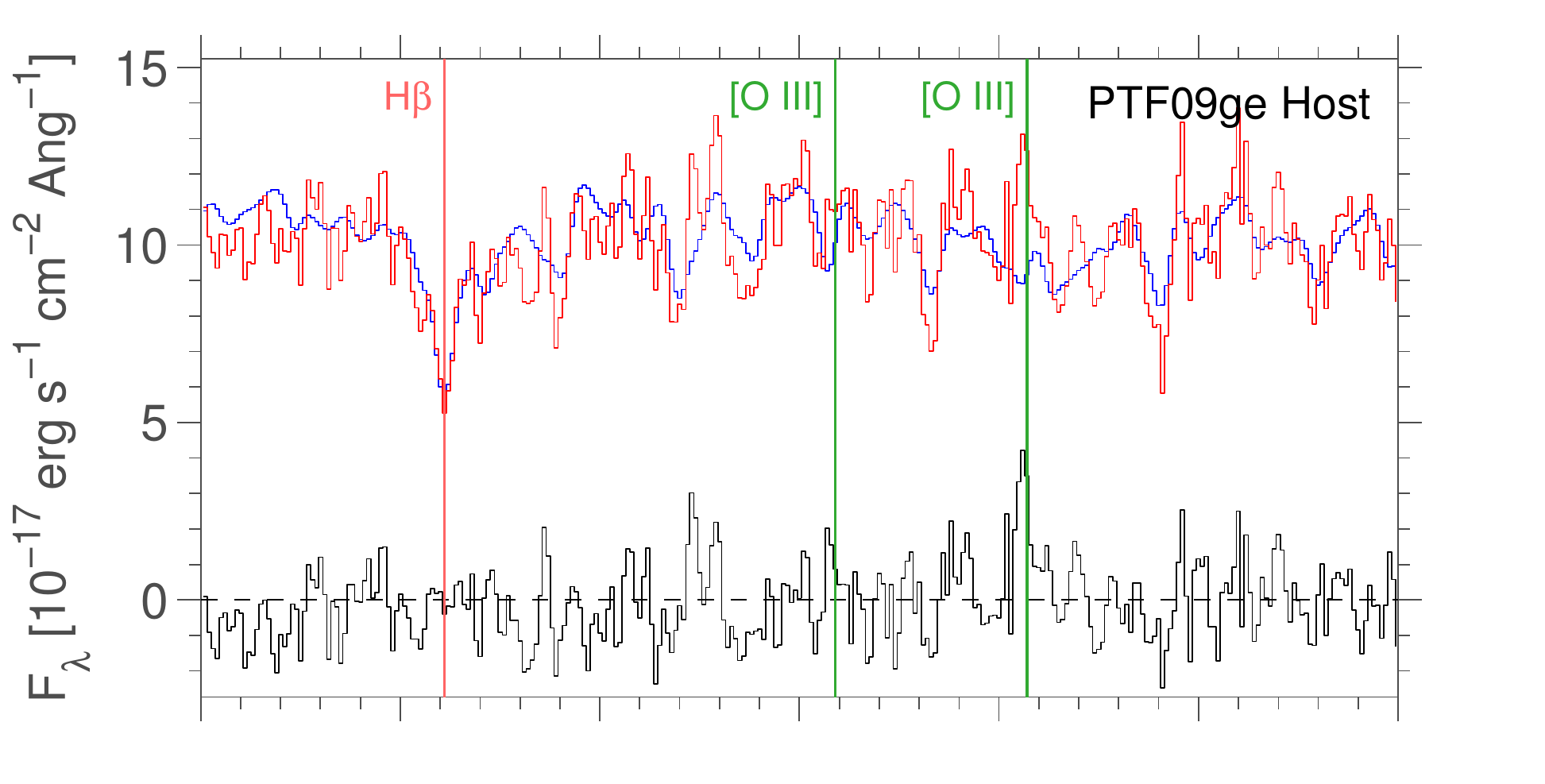}
\includegraphics[width=\columnwidth]{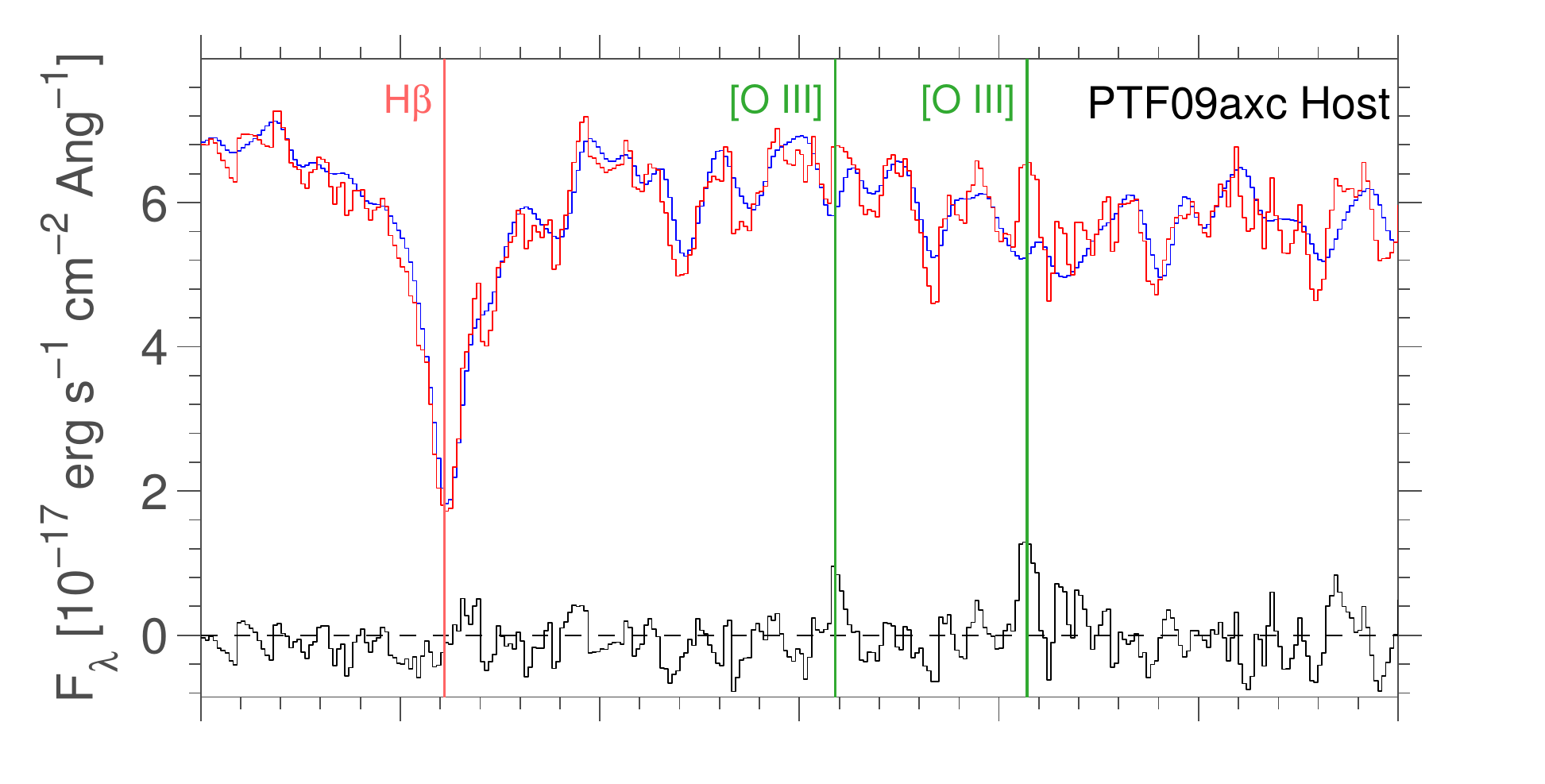}
\includegraphics[width=\columnwidth]{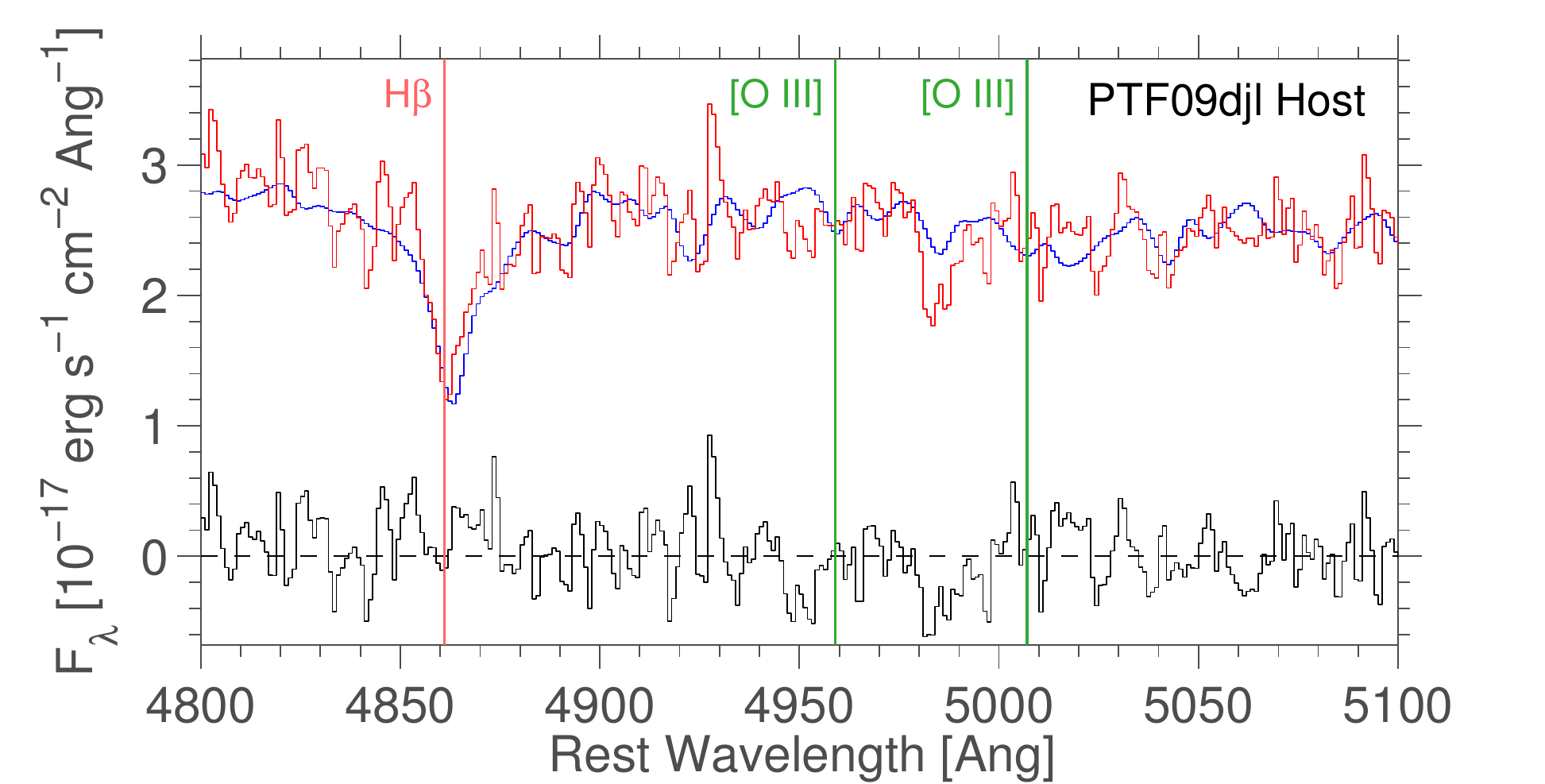}
\caption{\label{fig:sspptf}Host spectra (red), fitted stellar spectral templates (blue) and residuals (black). The spectra and templates are shifted in flux for clarity. PTF09axc shows an [OIII] $5007\,\textrm{\AA}$ emission line in the residual spectrum, marginally consistent with a weak AGN.}
\end{figure}

To evaluate the central emission more carefully, we subtract stellar spectral templates from the host-galaxy integrated spectra using the Bruzual \& Charlot (2003) stellar population synthesis models via the \textsc{starlight} code (Cid Fernandes et al. 2005). We find no signs of AGN emission lines in the host spectra of PTF09ge and PTF09djl, but for the host of PTF09axc we detect an [OIII] $5007\,\textrm{\AA}$ luminosity of $2.4\pm0.3\times10^{39}$ erg s$^{-1}$ (Fig. \ref{fig:sspptf}). We find an [OIII]/Hbeta ratio of $>3.4$, which indicates that the host galaxy of PTF09axc may contain a very weak AGN.

We find no X-ray sources in the {\it ROSAT} all sky survey at the positions of our targets between July 1990 and February 1991. Comparing to other {\it ROSAT} sources in the field, and using the Voges at el. (1999) conversions assuming a hardness ratio (HR1) $<0.5$, we derive $3{\sigma}$ upper limits for the X-ray flux of at the positions of PTF09ge, PTF09axc and PTF09djl as listed in Table \ref{tab:xray}. 

We obtained target-of-opportunity X-ray observations of the fields of PTF09ge, PTF09axc, and PTF09djl with the \textit{Swift} satellite (Gehrels et al. 2004) during March-April 2014 (i.e. five years after the outbursts). Data obtained by the on-board X-Ray Telescope (XRT; Burrows et al. 2005) was analyzed with the automated gamma-ray burst pipeline outlined in Evans et al. (2009). We find no X-ray flux at the positions of PTF09ge and PTF09djl (the derived limits listed Table \ref{tab:xray} assume a power law spectrum with a photon index of $2$), but we do detect X-ray emission at the position of PTF09axc corresponding to a luminosity of $7.13^{+12.22}_{-3.06}\times10^{42}$ erg s$^{-1}$. This luminosity, together with that in the host [OIII] $5007\,\textrm{\AA}$ line discussed above, is roughly consistent with AGNs on the low-luminosity end of the Heckman et al. (2005) local sample. The X-ray luminosity is likely too high (given the stellar mass of this galaxy) for an accreting-binary origin (Hornschemeier et al. 2005). We currently can not constrain any time variability in this X-ray signal, and therefore are not able to determine if it is related to PTF09axc directly.

\begin{table}
\caption{\label{tab:xray}{\it ROSAT} and {\it Swift} X-ray observations at the locations of our three nuclear transients. Two are not detected in either telescope ($3\sigma$ upper-limits shown). One (PTF09axc) is detected in the XRT data. This luminosity is marginally consistent with a weak AGN (Heckman et al. 2005) and too bright for an X-ray binary (Hornschemeier et al. 2005).}
\begin{tabular*}{\columnwidth}{l @{\extracolsep{\fill}} ll}
\hline
\hline
{Object} & {{\it ROSAT} $0.1$-$2.5$ keV} & {{\it Swift} XRT $0.3$-$10$ keV} \tabularnewline
{} & {(7/1990 - 2/1991)} & {(3/2014 - 4/2014)} \tabularnewline
{} & {[erg s$^{-1}$]} & {[erg s$^{-1}$]} \tabularnewline
\hline
{09ge} & {$<4.33\times10^{42}$} & {$<2.27\times10^{42}$} \tabularnewline
{09axc} & {$<1.86\times10^{43}$} & {$7.13^{+12.22}_{-3.06}\times10^{42}$} \tabularnewline
{09djl} & {$<6.28\times10^{43}$} & {$<1.91\times10^{43}$} \tabularnewline
\hline
\end{tabular*}
\end{table}

\begin{figure}
\includegraphics[width=\columnwidth]{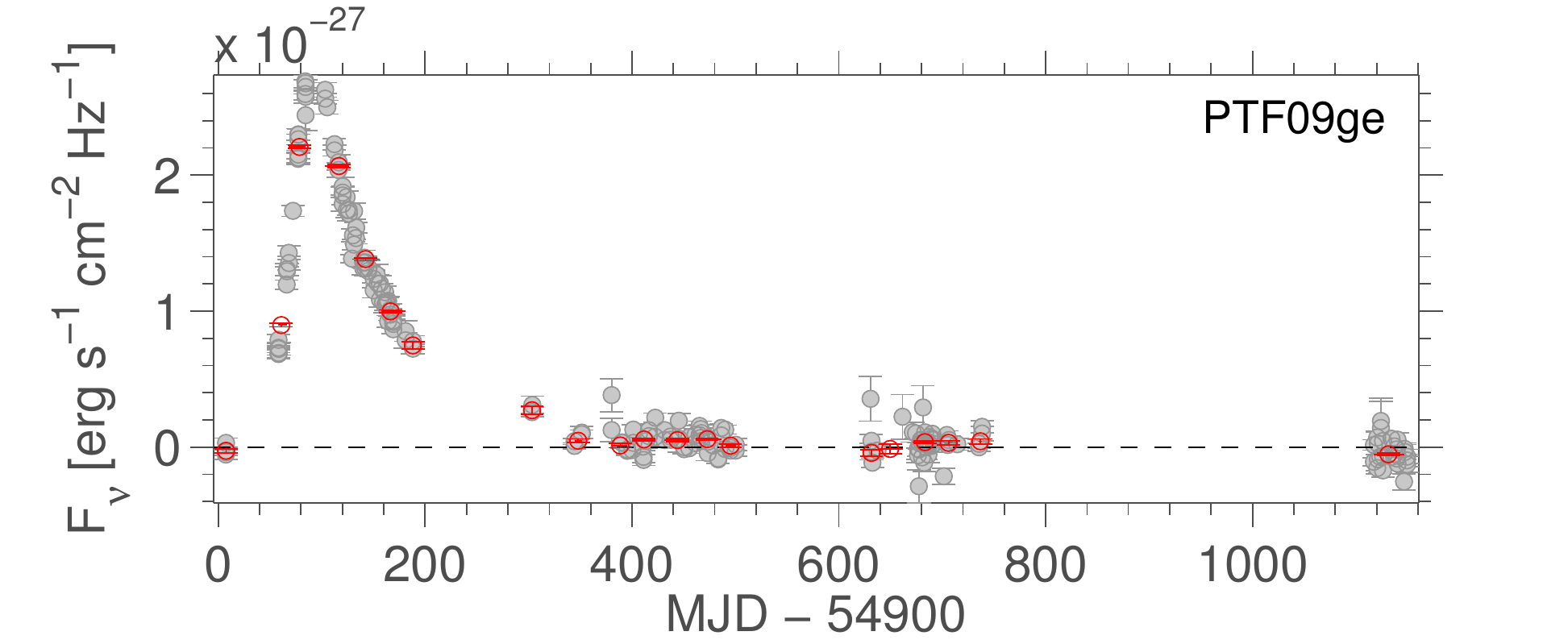}
\includegraphics[width=\columnwidth]{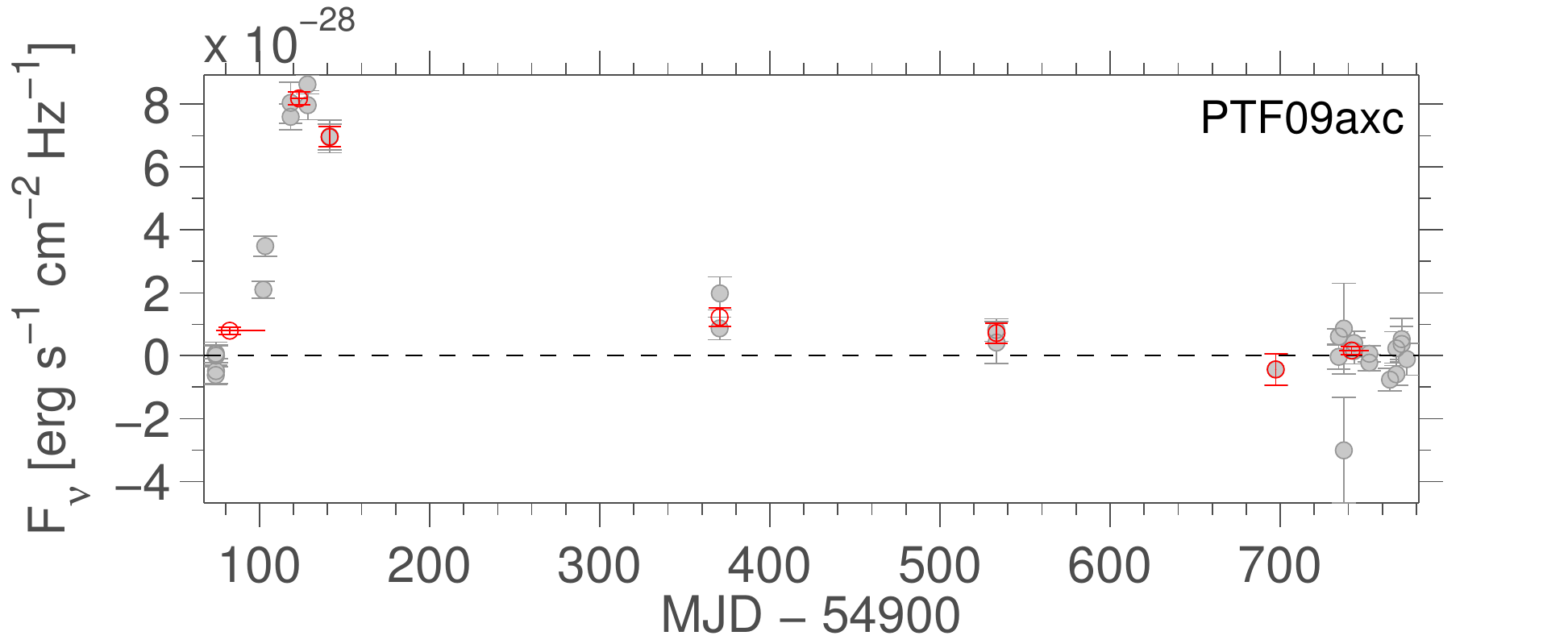}
\includegraphics[width=\columnwidth]{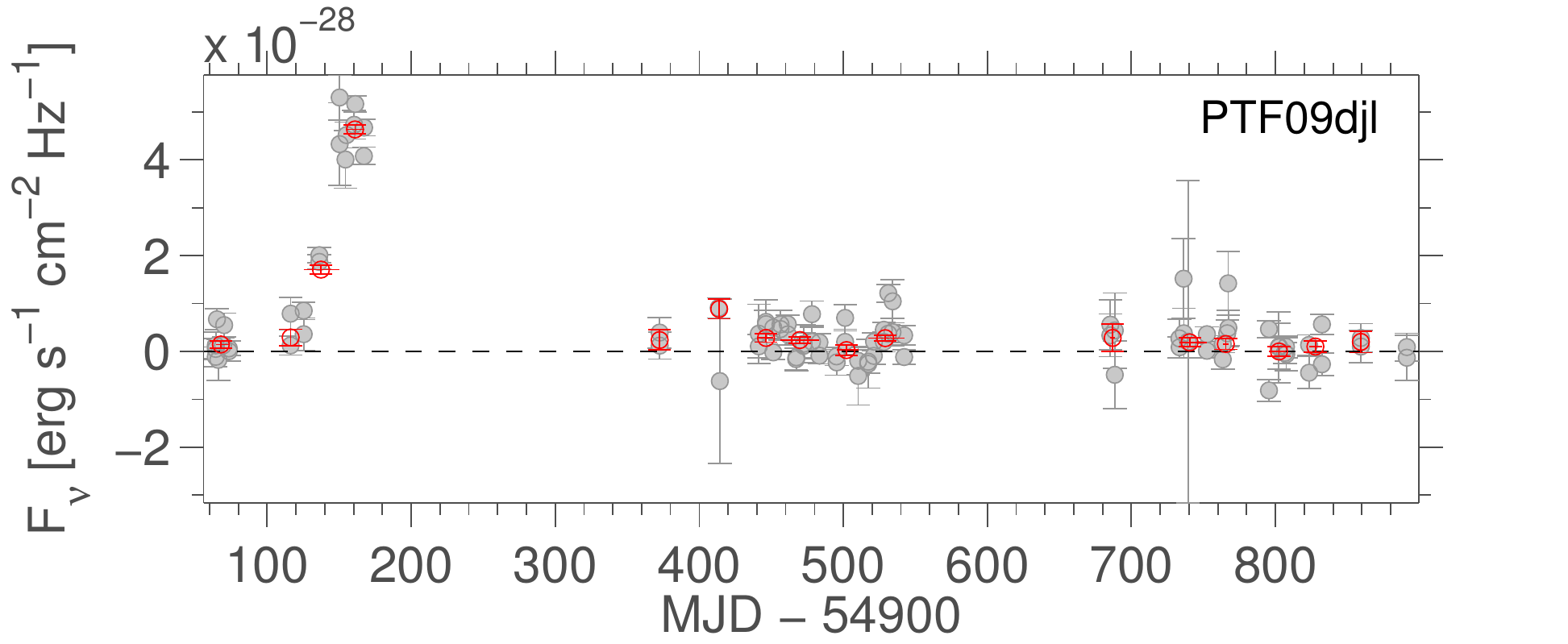}
\caption{\label{fig:longtermflux}Long term host-subtracted $R$-band P48 light curves showing no subsequent activity at the locations of PTF09ge, PTF09axc and PTF09djl for a few hundred days after the original outburst (grey: raw data; red: 30-day binned data).}
\end{figure}

Van Velzen et al. (2011) rejected a few of their TDE candidates as AGN based on photometric variability beyond the season containing the flare. Here we find no evidence for additional eruptions of our events during the years of available PTF coverage (Fig. \ref{fig:longtermflux}).

We conclude that these three outbursts are not likely due to AGNs, though the host of PTF09axc may also contain an extremely weak AGN. 

\subsection{A Sequence of H- to He-Rich Events}

\begin{figure}
\includegraphics[width=\columnwidth]{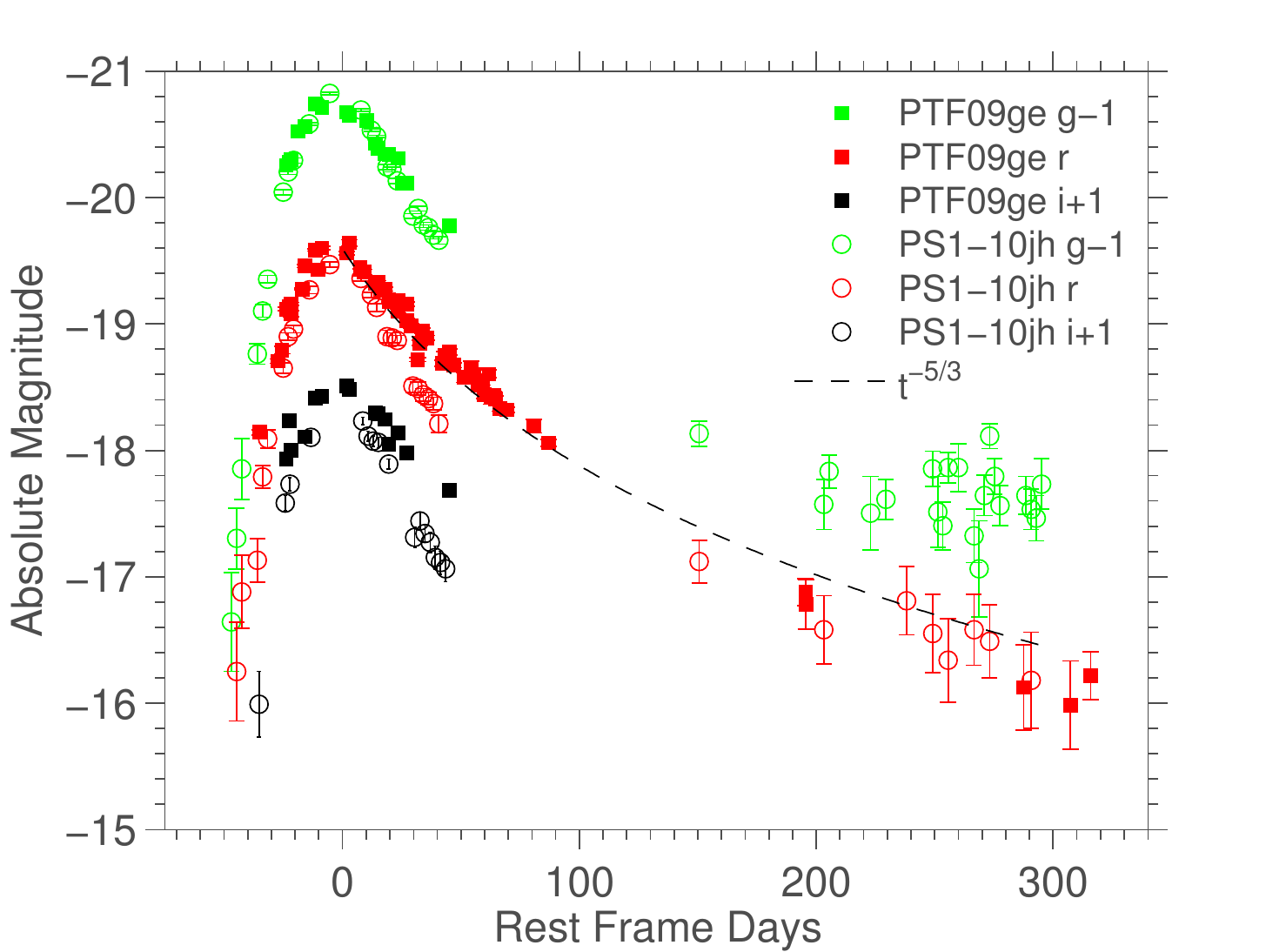}
\caption{\label{fig:09ge_phot}The $g$, $r$ and $i$-band light curves of PTF09ge from P48 and P60, compared to those of PS1-10jh (G12). Both events show very similar photometric behavior. A $t^{-5/3}$ decline rate is also shown.}
\end{figure}

\begin{figure}
\includegraphics[width=\columnwidth]{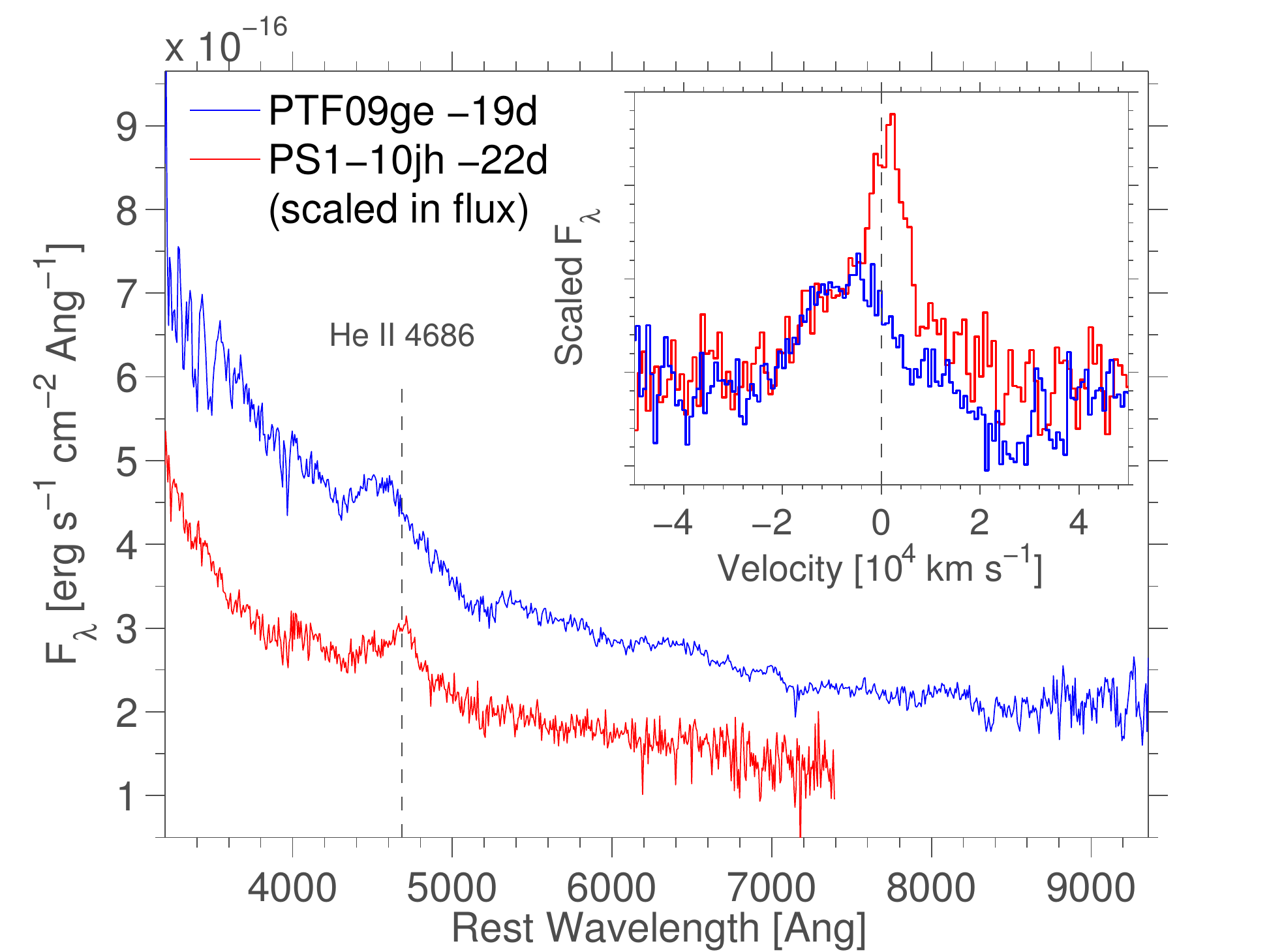}
\caption{\label{fig:09ge_spec}Spectra of PTF09ge and PS1-10jh (G12) at similar phases (shown in days relative to peak). The PS1-10jh spectrum is scaled in flux to the distance of PTF09ge. Both events show very similar spectral properties, namely broad He II $4686\,\textrm{\AA}$ on top of a blue continuum. \textsc{Inset}: The He II $4686\,{\textrm{\AA}}$ line profile of PTF09ge and PS1-10jh after host and continuum subtraction.}
\end{figure}

PTF09ge shows very similar photometric and spectroscopic behavior to PS1-10jh (Figs. \ref{fig:09ge_phot} and \ref{fig:09ge_spec}), identified by G12 as a likely He-rich TDE. Both objects display broad He II emission superimposed on a blue continuum. We find a $\sim1000$ km~s$^{-1}$ blueshift in the He II $4686\,{\textrm{\AA}}$ emission peak of PTF09ge. A similar blue ``wing'' was observed in PS1-10jh (see inset of Figure \ref{fig:09ge_spec}), suggesting that the PS1-10jh line profile is made of two components: an extended blueshifted component (seen also in PTF09ge), and an intermediate-width component (not seen in PTF09ge, but apparent also in the H$\alpha$ profile of ASASSN-14ae; Holoien et al. 2014). For PTF09ge, we further identify a possible broad absorption feature redshifted by $\sim3000$ km~s$^{-1}$, but this could be due to remaining narrow Fe II $5018\,\textrm{\AA}$ and $5169\,\textrm{\AA}$ contamination from the host.

The single-band, sparsely sampled light curves of PTF09axc and PTF09djl show similar rise-times to PTF09ge (Fig. \ref{fig:all_phot}) but their spectra exhibit broad H$\alpha$ emission (Fig. \ref{fig:all_specs}). Van Velzen et al. (2011) noted possible hydrogen emission in their TDE2 as well, albeit narrower and less prominent than in our events. 

\begin{figure}
\includegraphics[width=\columnwidth]{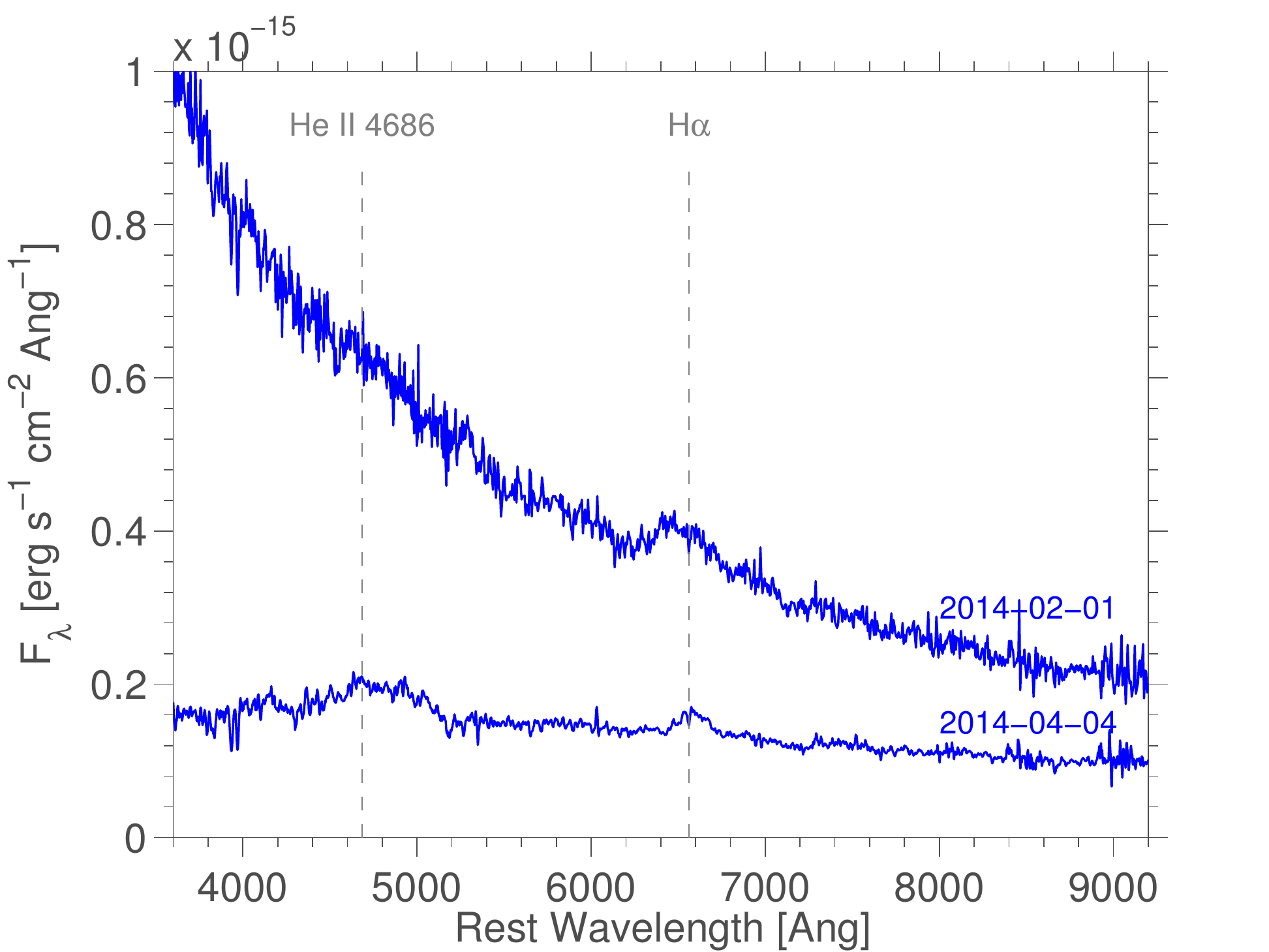}
\caption{\label{fig:aivspec}Spectra of ASASSN-14ae. The early spectrum displays broad H${\alpha}$ on top of a blue continuum. An additional broad component at He II $4686\,\textrm{\AA}$ appears in the later (cooler) spectrum.\\}
\end{figure}

Following the announcement of ASASSN-14ae by Prieto et al. (2014), and later discovery by iPTF, we obtained spectra of it with DBSP on P200 on 2014-Feb-01 and 2014-Apr-04 (Fig. \ref{fig:aivspec}). The later spectrum shows emission lines of both H and He II, similar to those noted for SDSS J0748 by Wang et al. (2011).

\begin{figure*}
\centering
\includegraphics[trim=0 0 40 0, clip,height=11cm]{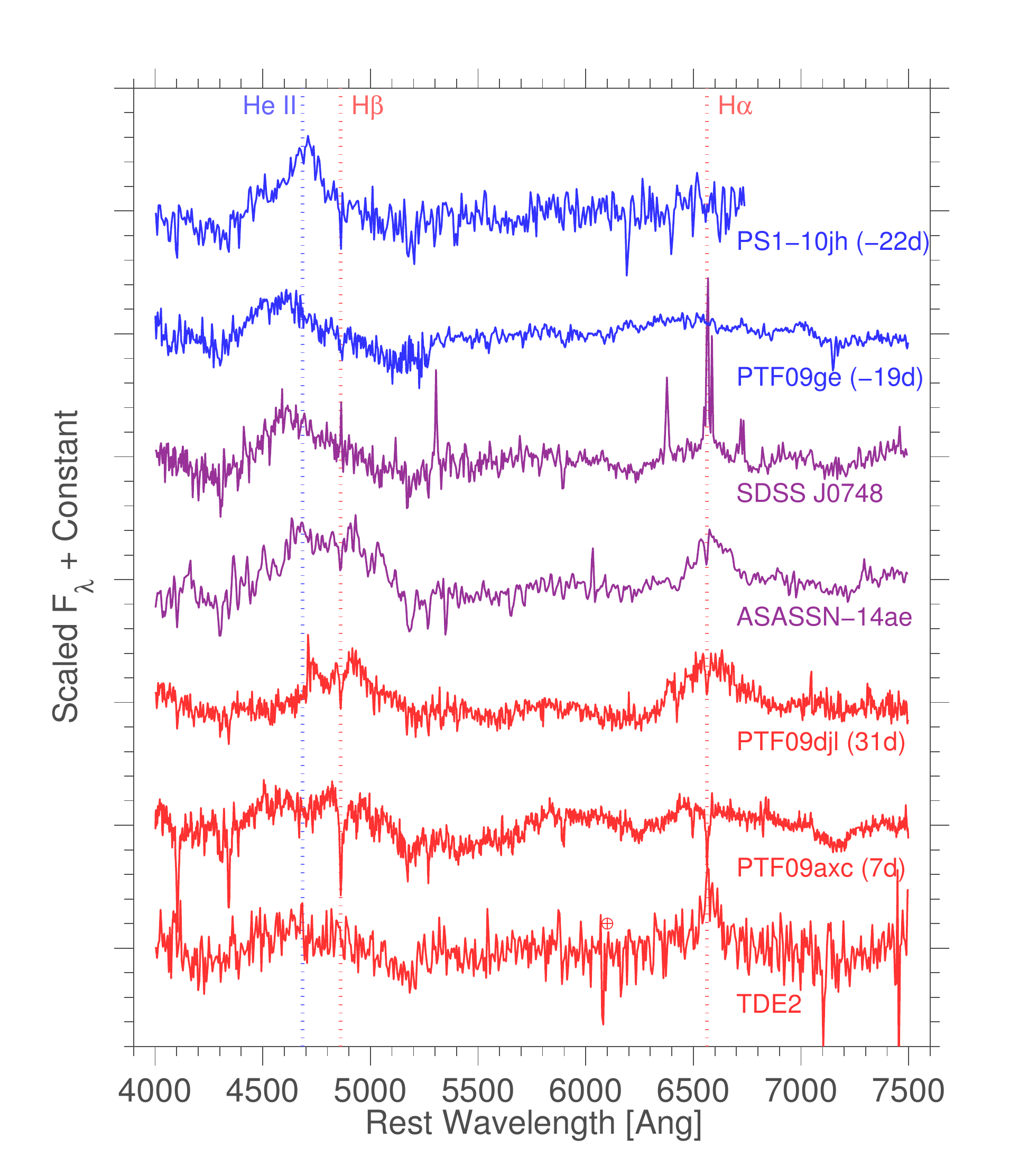}\includegraphics[trim=10 0 10 0, clip,height=11cm]{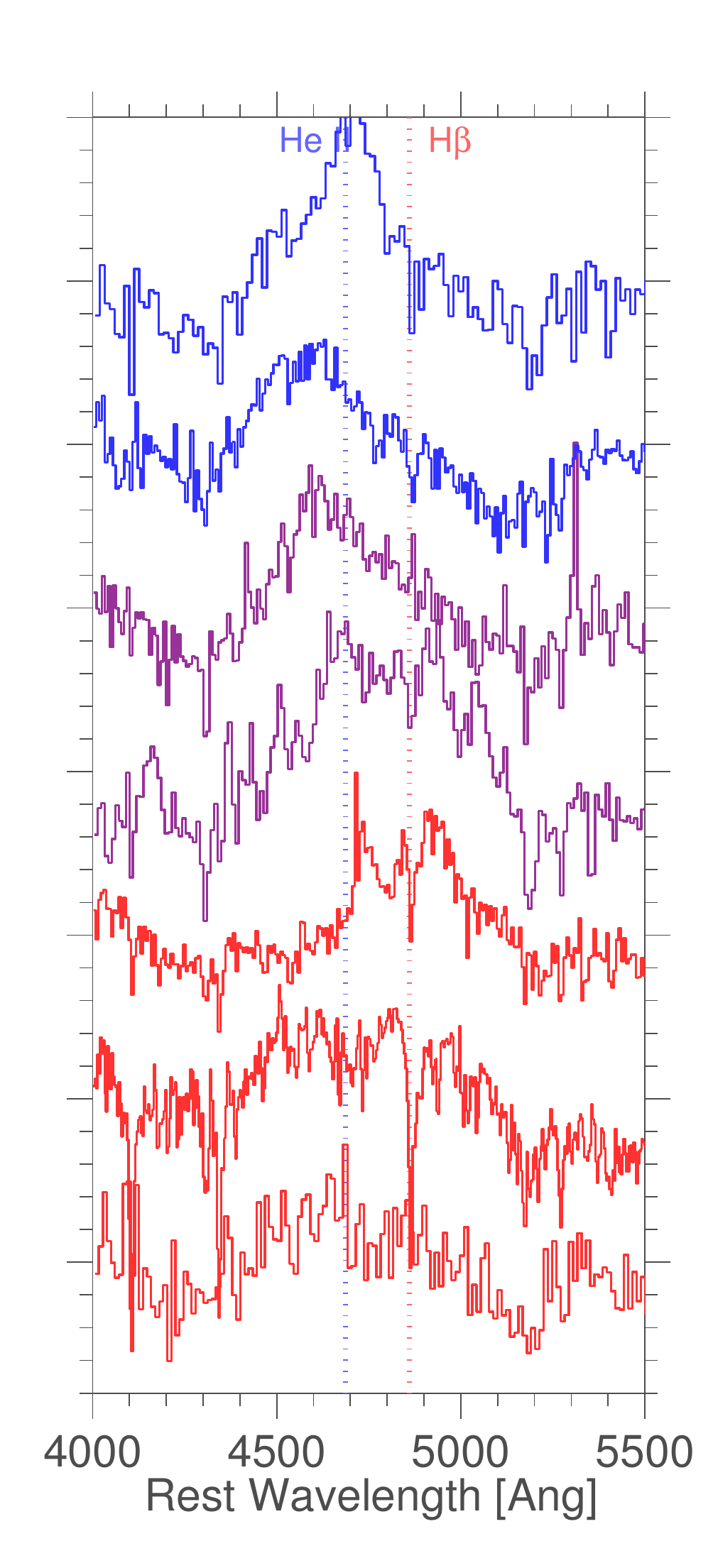}\includegraphics[trim=10 0 00 0, clip,height=11cm]{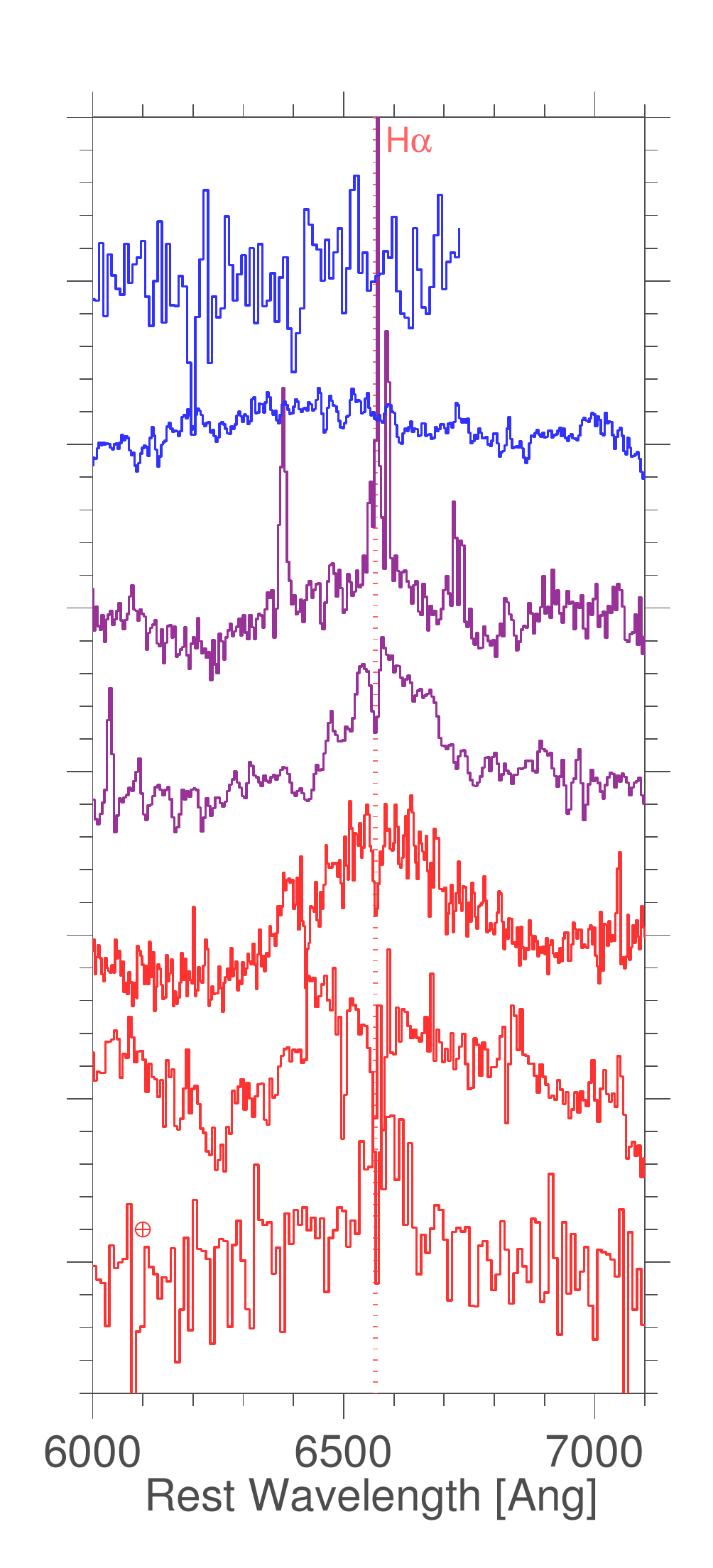}
\caption{\label{fig:hhe}Continuum-subtracted spectra of our three PTF TDE candidates together with PS1-10jh (G12), ASASSN-14ae, SDSS J0748 (Wang et al. 2011) and TDE2 (van Velzen et al. 2011). Phases are shown relative to peak. A progression from He-rich to H-rich events is apparent. The middle and right panels present more detailed views of the regions around the marked lines.}
\end{figure*}

In total, these seven TDE candidates span a continuous sequence of spectral types, from He-dominated (PS1-10jh, PTF09ge) to H-dominated (PTF09djl, PTF09axc and possibly TDE2) through intermediate H+He events (SDSS J0748, ASASN-14ae). The spectra of these events have different continuum shapes (either due to intrinsic differences between the events, or because of extinction differences). To isolate the differences in the emission features, we remove a $2^{nd}$ order polynomial fit from each spectrum and present the continuum-subtracted spectra of this sample in Figure \ref{fig:hhe}. 

\begin{figure*}
\centering
\includegraphics[trim=0 0 30 0, clip,height=11cm]{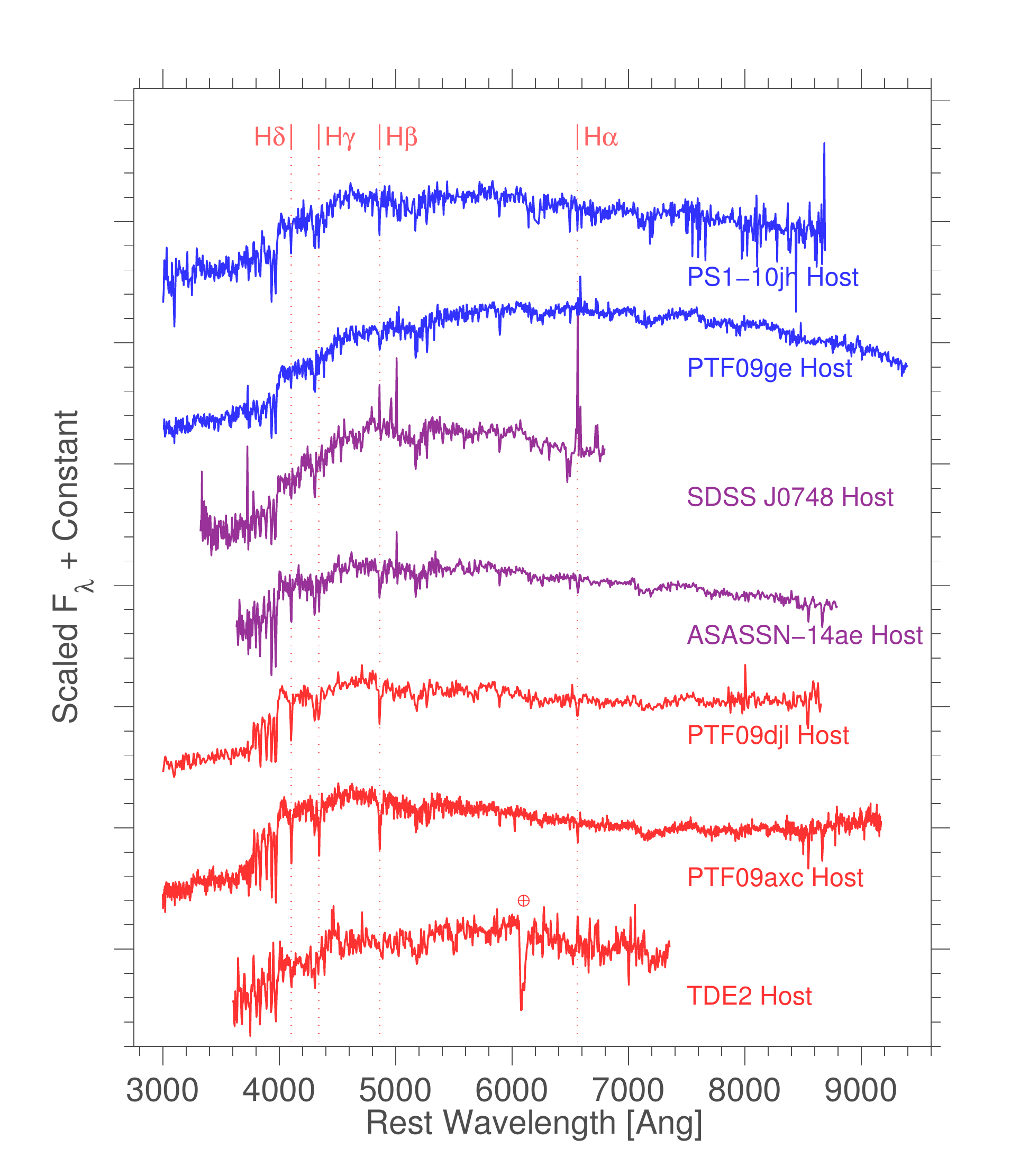}\includegraphics[trim=10 0 0 0, clip,height=11cm]{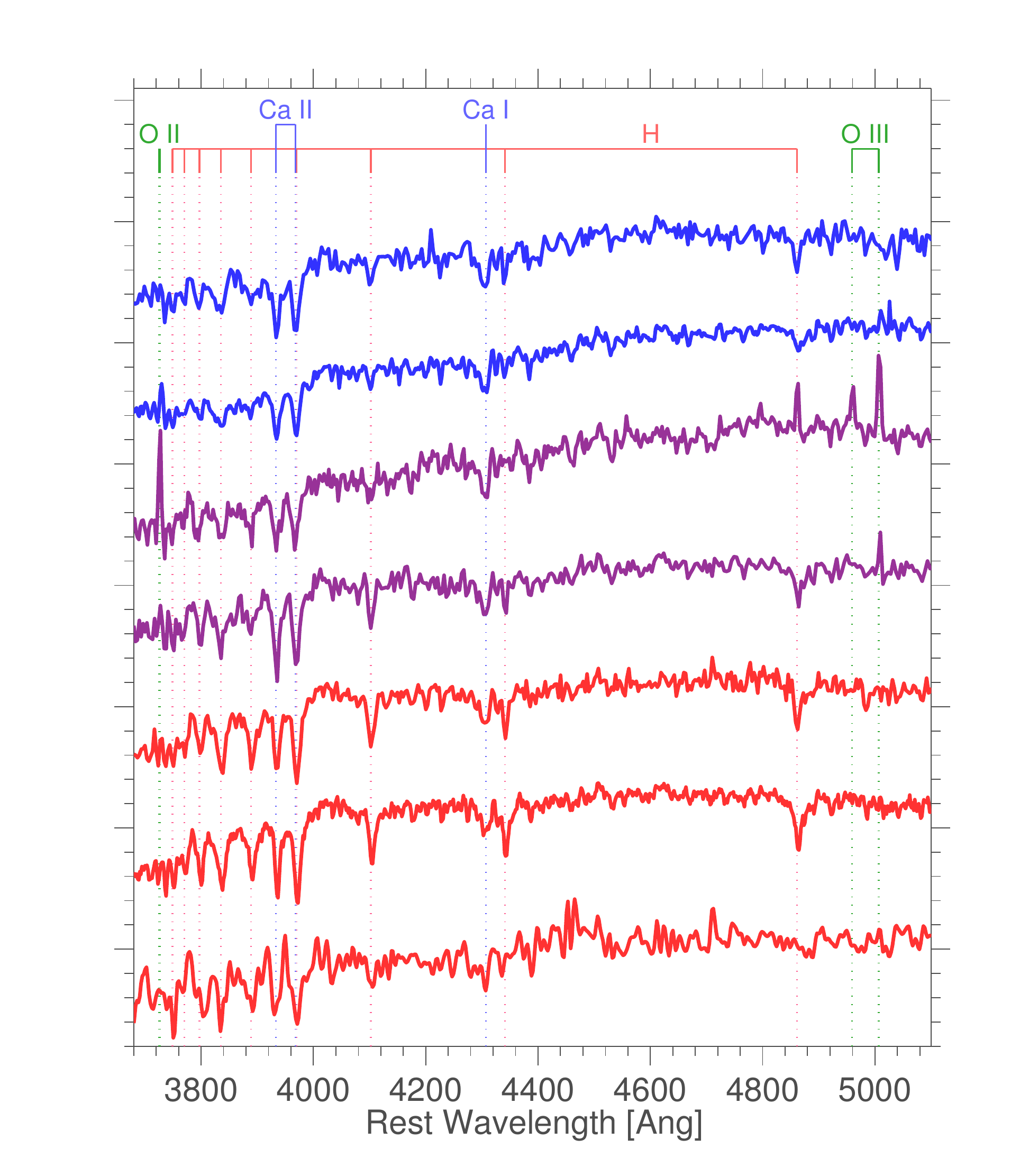}
\caption{\label{fig:hhehost}Host galaxy spectra of the TDE candidates from Figure \ref{fig:hhe} (the spectrum of SDSS J0748 is from Yang et al. 2013; the spectrum of the ASASSN-14ae host is from SDSS DR10). Most show similar signs of low ongoing star formation. Balmer-series absorption lines identify some of these hosts as rare E+A galaxies. The right panel presents a more detailed view of the marked lines.}
\end{figure*}

The He-dominated spectra in Figure \ref{fig:hhe} are taken at earlier phases compared to the rest, suggesting the features observed could be time-dependent. However, PS1-10jh remained He-dominated out to 254 days post-peak (G12), while ASASSN-14ae displays H from its first spectrum (Holoien et al. 2014; Fig. \ref{fig:aivspec}).

\subsection{\label{hosts2}Host Galaxies}

We present host galaxy spectra for these seven TDE candidates in Figure \ref{fig:hhehost}. The properties of the hosts of the PTF events were discussed in Section \ref{hosts}. Here we add our own spectrum of the host galaxy of PS1-10jh (obtained with LRIS on Keck I on April 29, 2014), the TDE2 post-flare spectrum from van Velzen et al. (2011), the SDSS J0742 post-flare spectrum from Yang et al. (2013), and the ASASN-14ae pre-flare spectrum from SDSS DR10. All host galaxies show blue Balmer-sequence absorption features, typical of E+A galaxies, though some display H${\alpha}$, [O II] and [O III] in emission.

\begin{figure}
\includegraphics[width=\columnwidth]{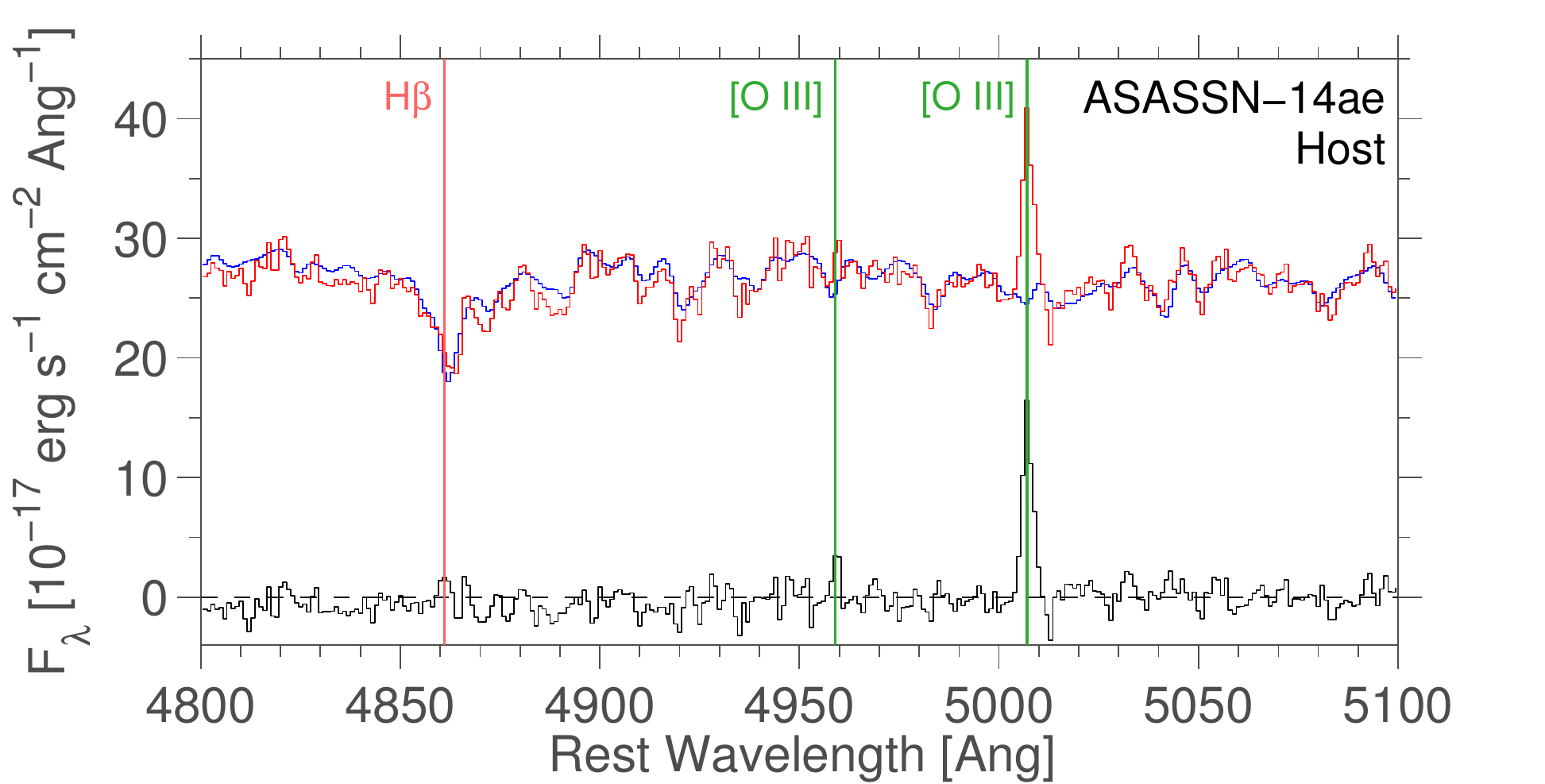}
\includegraphics[width=\columnwidth]{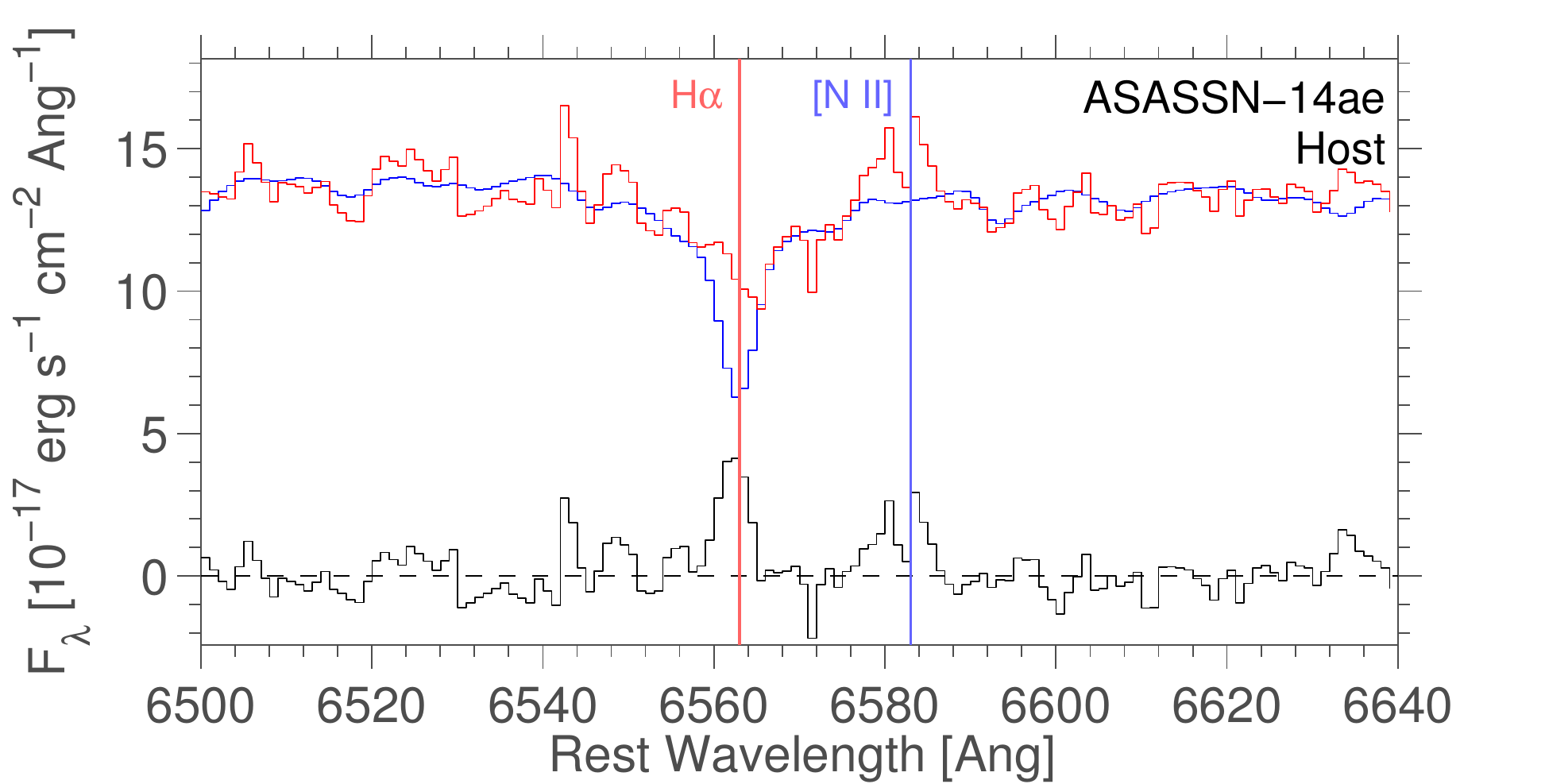}
\caption{\label{fig:ssptde}Same as Figure \ref{fig:sspptf} but for the host galaxy of ASASSN-14ae. The [OIII]/H$\beta$ ratio and the [OIII] luminosity are consistent with a weak AGN.}
\end{figure}

The SDSS host-galaxy spectrum of ASASSN-14ae, in particular, shows prominent [OIII] $5007\,\textrm{\AA}$ emission. We perform the same analysis described in section \ref{sec:agn} to extract the non-stellar component of the host spectrum (Fig. \ref{fig:ssptde}). We detect [OIII] $5007\,\textrm{\AA}$ and H$\alpha$ at $5.2\times10^{-16}$ and $1.7\times10^{-16}$ ergs s$^{-1}$ cm$^{-2}$ respectively in the stellar-subtracted spectrum, and find no emission from H$\beta$ and [OII] $3727\,\textrm{\AA}$ to $3\sigma$ upper limits of $7\times10^{-17}$ and $3\times10^{-17}$ ergs s$^{-1}$ cm$^{-2}$, respectively. For [NII] $6583\,\textrm{\AA}$, contamination from sky emission make it difficult to determine the existence of the line and we set a conservative $3\sigma$ upper limit of $10^{-16}$ ergs s$^{-1}$ cm$^{-2}$ for it. The measured [OIII]/H$\beta$ and [NII]/H$\alpha$ ratios are $>7.4$ and $<0.59$, respectively. Such a high [OIII]/H$\beta$ ratio suggests the presence of central accreting black hole (Baldwin, Philips \& Terlevich 1981).  The low luminosity of [OIII] $5007\,\textrm{\AA}$ (at $2.3\times10^{39}$ ergs s$^{-1}$) is consistent with the host containing a weak AGN. 

We repeat the analysis of Section \ref{hosts} for the hosts of these TDE candidates, and present the deduced parameters in Table \ref{tab:tdehost_params} together with those of the PTF TDE candidates from Table \ref{tab:host_params} for easy comparison. We translate our measured stellar masses to bulge masses through the Gadotti (2009) conversion using the SDSS R90/R50 ratios\footnote{R90 and R50 are the radii enclosing $90\%$ and $50\%$ of the galaxy light respectively.} in the $r$-band. We then use the relation of H{\"a}ring \& Rix (2004) to translate the bulge mass to the SMBH mass. The R90 and R50 values for the host galaxy of PS1-10jh have very large errors due to the low luminosity of this galaxy. We therefore adopt the SMBH mass obtained by G12 in their scaling from the host galaxy mass. All SMBH masses are presented in Table \ref{tab:tdehost_params}.

\renewcommand{\arraystretch}{1.5}
\begin{table*}
\caption{\label{tab:tdehost_params}Same as Table \ref{tab:host_params} for the host galaxies of the PTF TDE candidates together with the additional TDE candidates presented in Figure \ref{fig:hhe}. We add the SMBH mass calculated using the Gadotti (2009) and H{\"a}ring \& Rix (2004) relations (for PS1-10jh we adopt the value from G12).}
\begin{tabular*}{\textwidth}{lllllllll}
\hline
\hline
{Host} & \multicolumn{4}{l}{Photometric Analysis} & \multicolumn{4}{l}{Spectroscopic Analysis}\tabularnewline
{} &  {M}  & {$\textrm{M}_\textrm{BH}$} & {SFR} & {sSFR} & {SFR} & {$12+{\log}$(O/H)} & {[M/H]} & {Age} \tabularnewline
{} &  {[$10^{10}M_{\odot}$]} & {[$10^6M_\odot$]} & {[$M_{\odot}\textrm{yr}^{-1}$]} & {[$10^{-10}\textrm{yr}^{-1}$]} & {[$M_{\odot}\textrm{yr}^{-1}$]} & {} & {} & {[Gyr]}\tabularnewline
\hline
{PTF09ge} & {$1.05$ ($1.03$, $1.35$)} & {$5.65_{-0.98}^{+3.02}$} & {n/a} & {n/a} & {$0.10$ ($0.05$)} & {$8.873$ ($0.064$)} & {$-0.196$} & {$7.035$}\tabularnewline
{PTF09axc} & {$1.23$ ($1.16$, $1.28$)} & {$2.69_{-0.64}^{+0.66}$} & {$<16.11$} & {$<1.31$} & {$0.04$ ($0.02$)} & {n/a} & {$-0.356$} & {$4.469$}\tabularnewline
{PTF09djl} & {$1.86$ ($1.07$, $3.73$)} & {$3.57_{-2.96}^{+9.97}$} & {$3.42$ ($1.22$, $4.19$)} & {$1.84$ ($0.34$, $3.02$)} & {n/a} & {n/a} & {$-0.218$} & {$4.461$}\tabularnewline
{PS1-10jh} & {$0.67$ ($0.48$, $1.12$)} & {$4_{-2}^{+4}$} & {$1.21$ ($0.29$, $1.28$)} & {$1.81$ ($0.30$, $2.02$)} & {n/a} & {n/a} & {$-0.215$} & {$5.599$}\tabularnewline
{SDSS J0748} & {$3.40$ ($2.79$, $3.57$)} & {$11.78_{-3.56}^{+2.29}$} & {$2.76$ ($2.58$, $3.58$)} & {$0.81$ ($0.76$, $1.22$)} & {$0.35$ ($0.40$)} & {$8.760$ ($1.723$)} & {$-0.199$} & {$6.265$}\tabularnewline
{ASASSN-14ae} & {$0.60$ ($0.52$, $0.8$)} & {$2.45_{-0.74}^{+1.55}$} & {n/a} & {n/a} & {$0.02$ ($0.01$)} & {n/a} & {$-0.407$} & {$5.544$}\tabularnewline
{TDE2} & {$9.33$ ($8.43$, $13.06$)} & {$35.52_{-25.80}^{+55.31}$} & {$1.01$ ($0.53$, $1.22$)} & {$0.11$ ($0.04$, $0.13$)} & {n/a} & {n/a} & {$-0.319$} & {$4.627$}\tabularnewline
\hline
\end{tabular*}
\end{table*}

\renewcommand{\arraystretch}{1}

\subsection{Line Widths}

The velocity widths of the broad emission lines seen during a TDE could be indicative of the region from which the lines are being emitted. 

Bound material in a Keplerian orbit at the tidal radius $R_T$, will have a (circular) velocity
\begin{equation}
v_{T}\approx43700\left(\frac{M_{BH}}{10^{6}M_{\odot}}\right)^{1/3}\left(\frac{\rho_{*}}{\rho_{\odot}}\right)^{1/6}\textrm{km s}^{-1}
\end{equation}
(where $\rho_{*}$ is the average density of the disrupted star). Assuming $M_{BH}\propto\sigma_G^{\alpha}$ (where $\sigma_G$ is the velocity dispersion of the galaxy), and $L\propto\sigma_G^4$ (from the Faber-Jackson relation; Faber \& Jackson, 1976), it follows that: 
\begin{equation}
v_T{\propto}\left(L_{\rm host}\right)^{\alpha/12}\rho_{*}^{1/6}
\end{equation}
For typical values of ${\alpha}$ (e.g. $\alpha=4.42$; Kormendy \& Ho 2013), the stellar density will have a small influence and a correlation between the measured TDE line-widths and the host galaxy luminosity is expected, assuming the measured velocity indeed represents $v_T$. Guillochon et al. (2014), however, show that bound material could extend to larger radii than the standard truncated disk model assumes. Therefore bound material could have much lower velocities than $v_T$.

Another option for the origin of the emission lines is the outflowing unbound material. Strubbe \& Quataert (2009) find that the most energetic unbound material will move at velocities of approximately
\begin{equation}
\label{eq:unbound_v}
7500\left(\frac{R_T}{R_P}\right)\left(\frac{M_{BH}}{10^{6}M_{\odot}}\right)^{\frac{1}{6}}\left(\frac{M_{*}}{M_{\odot}}\right)^{\frac{1}{3}}\left(\frac{R_{*}}{R_{\odot}}\right)^{-\frac{1}{2}}\textrm{km s}^{-1}
\end{equation}
For main sequence stars, the stellar mass and radius nearly cancel, but the $R_T/R_P$ factor here may smear any remaining correlation.

Strubbe \& Quataert (2009) note that lines from the outflowing material alone will be bulk-blueshifted or redshifted. It is therefore possible that a broadening of the lines would be caused by a combination of emission from bound and the outflowing material.

We fit a Gaussian function to each broad emission feature in continuum-subtracted spectra of the TDE candidates presented in Figure \ref{fig:hhe} and use it to estimate the velocity width of these lines. We change the continuum subtraction parameters and take the scatter in the fitted width as its error. Our measured $1\sigma$ line widths are presented in Table \ref{tab:vm} (our values for the line widths of TDE2 and PS1-10jh are consisted with those reported by van Velzen et al. 2011 and G12 respectively, though they quote full-width at half-maximum values while we use $1{\sigma}$).

The velocities are lower than the expected $v_{T}$ (at the tidal radius), and correspond to bound (circular) orbits at distances of $\sim20-80\left(M_{BH}/10^6M_{\odot}\right)^{2/3}\left(\rho_*/\rho_{\odot}\right)^{1/3}$ tidal radii from the SMBH, or to outflowing unbound trajectories.

\begin{table}
\caption{\label{tab:vm}Measured $1\sigma$ line widths of the H or He emission feature. Phases are shown relative to peak brightness. For PTF09djl, the average velocity from the three spectra is shown, as we see no evolution in the width of the main component within the errors.}
\begin{tabular*}{\columnwidth}{l @{\extracolsep{\fill}} llll}
\hline
\hline
{Object} & {Line} & {Phase} & {Line} \tabularnewline
{} & {} & {[days]} & {Width [km s$^{-1}$]} \tabularnewline
\hline
{PS1-10jh} & {He II $4686\,\textrm{\AA}$} & {$-22$} & \multicolumn{1}{r}{$5430\pm1460$} \tabularnewline
{PTF09ge} & {He II $4686\,\textrm{\AA}$} & {$-19$} & \multicolumn{1}{r}{$10070\pm670$} \tabularnewline
{SDSS J0748} & {He II $4686\,\textrm{\AA}$} & {n/a} & \multicolumn{1}{r}{$9950\pm510$} \tabularnewline
{ASASSN-14ae} & {H${\alpha}$} & {n/a} & \multicolumn{1}{r}{$3600\pm175$} \tabularnewline
{PTF09axc} & {H${\alpha}$} & {$7$} & \multicolumn{1}{r}{$11890\pm220$} \tabularnewline
{PTF09djl} & {H${\alpha}$} & {$2-62$} & \multicolumn{1}{r}{$6530\pm350$} \tabularnewline
{TDE2} & {H${\alpha}$} & {n/a} & \multicolumn{1}{r}{$3440\pm1100$} \tabularnewline
\hline
\end{tabular*}
\end{table}

\begin{figure}
\includegraphics[width=\columnwidth]{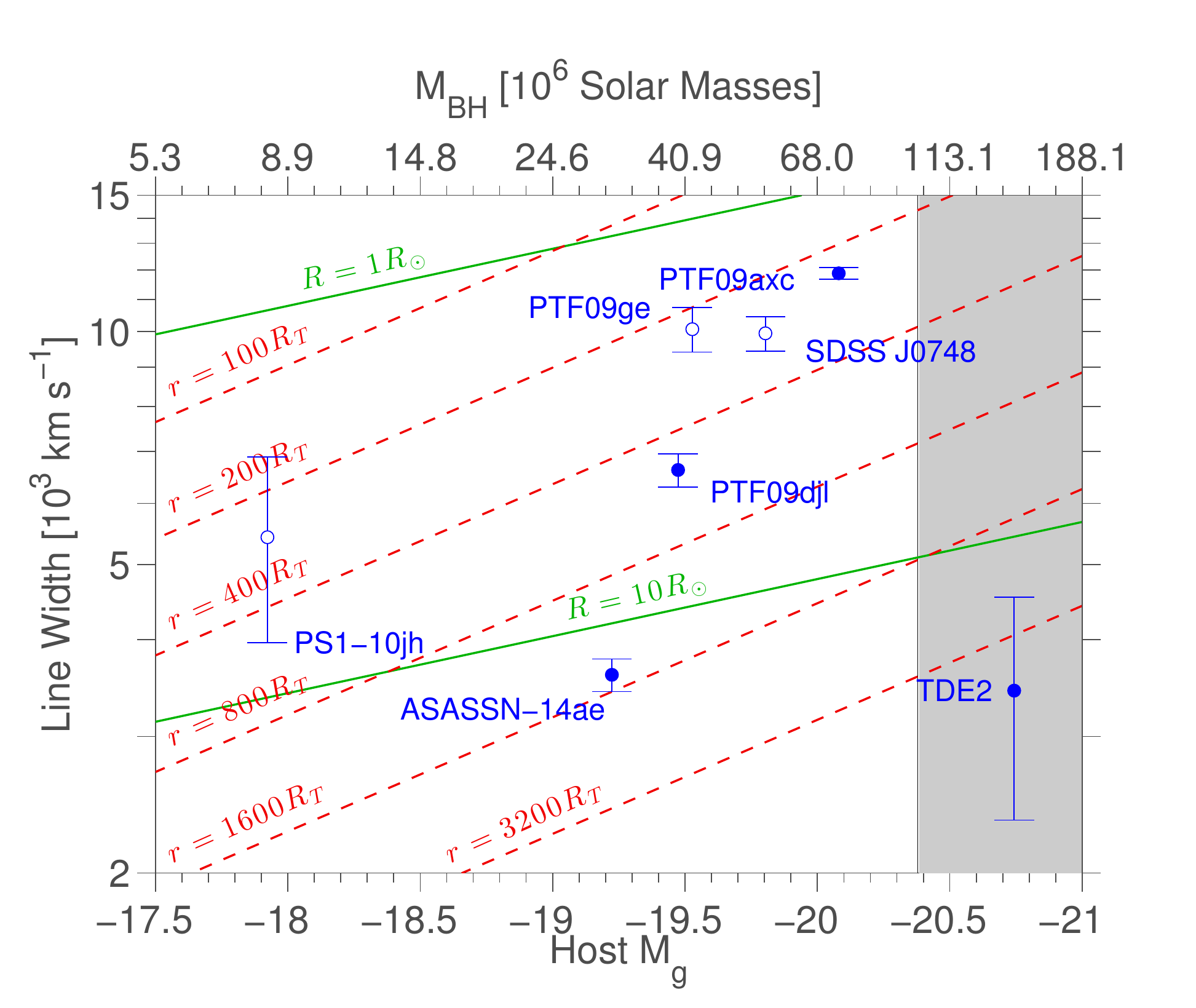}
\includegraphics[width=\columnwidth]{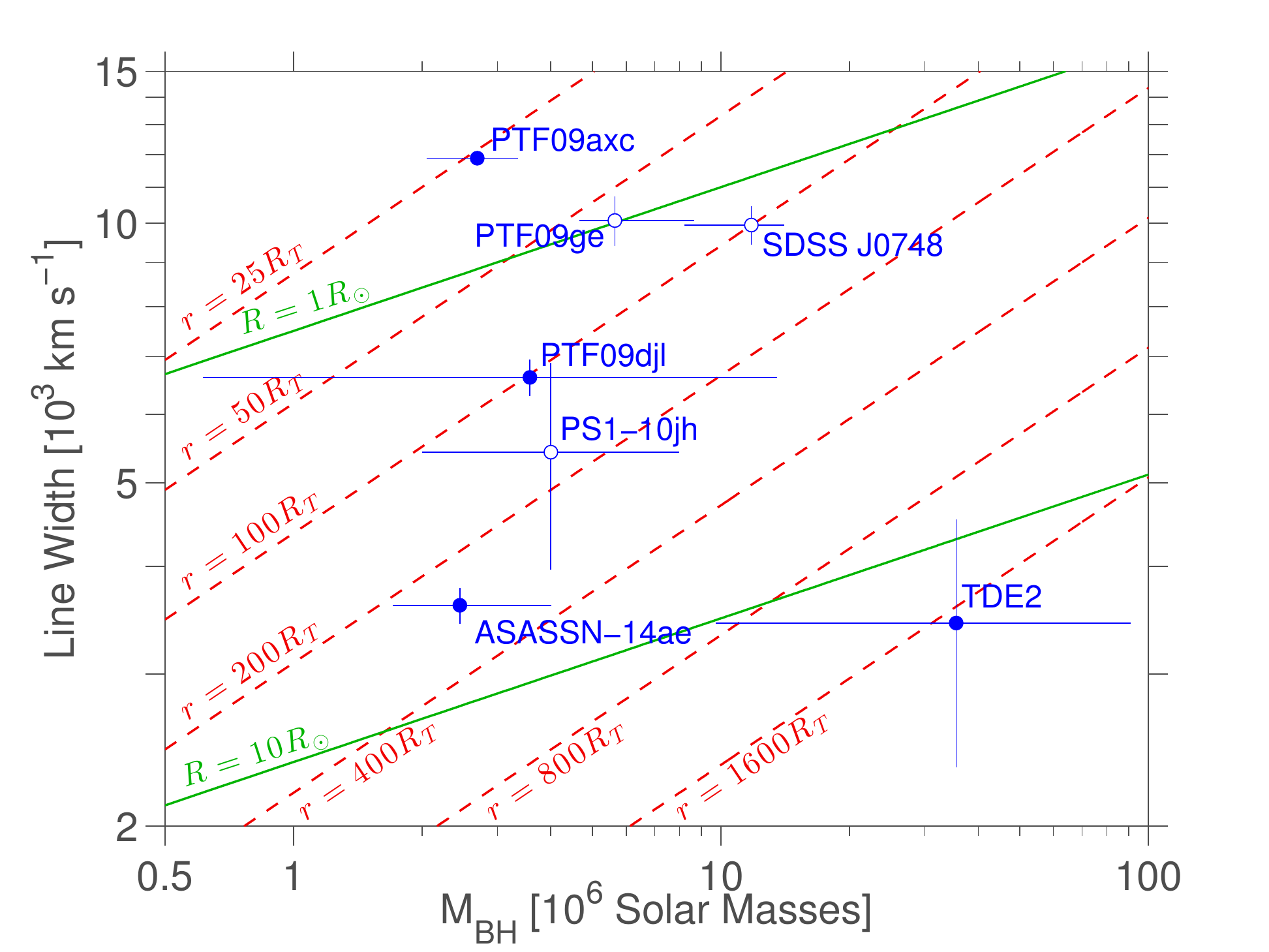}
\caption{\label{fig:mvcorr}TDE-candidate line widths of H (filled circles) and He (empty circles) emission lines vs. host galaxy K-corrected $g$-band absolute magnitudes (top) and SED-derived SMBH masses (bottom). The dashed red lines represent the expected velocity correlations for bound material at different radii (assuming a sun-like star). The solid green lines are the Strubbe \& Quatert (2009) velocities for the most energetic outflowing material assuming $R_P=R_T$ for a sun-like star or red giant as noted. For the top panel, we use the Bernardi et al. (2003) coefficients for the Faber-Jackson relation to derive these lines and to derive the top x-axis values from the host galaxy $g$-band absolute magnitudes. For the bottom panel, the SMBH masses are taken from Table \ref{tab:tdehost_params} (i.e. calculated using the Gadotti (2009) and H{\"a}ring \& Rix (2004) relations from our derived host stellar masses). The shaded region in the top panel denotes SMBH masses above $10^8M_\odot$, for which a TDE is not expected to be observed (assuming a sun-like star and a non-rotating SMBH).}
\end{figure}

We plot the observed line velocities vs. host galaxy magnitudes and vs. derived SMBH masses (Figure \ref{fig:mvcorr}, top and bottom panels respectively). For the top panel we use K-corrected (Chilingarian et al. 2010)\footnote{Obtained using g-r colors through the ``K-corrections calculator'' at http://kcor.sai.msu.ru/} host galaxy $g$-band magnitudes. We translate these magnitudes to SMBH mass using the Bernardi et al. (2003) coefficients for the Faber-Jackson relation and assuming the $M-\sigma$ relation found by Kormendy \& Ho (2013). For the bottom panel we calculate the SMBH masses from our derived galaxy stellar masses using the Gadotti (2009) and H{\"a}ring \& Rix (2004) relations (see Section \ref{hosts2}). In both panels we plot the expected correlations for bound material at several times the tidal radius for a sun-like disrupted star, and for unbound material with $R_T=R_P$ (Strubbe \& Quataert 2009). The overall scale of the velocities can be seen to be consistent with large bound radii or unbound velocities (as stated by G12 for PS1-10jh). 

There does not seem to be a robust correlation of the line widths with either host galaxy magnitude or SMBH mass, encompassing all events. This suggests that the simple association of velocities to bound circular Keplerian orbits is likely incorrect. Additional outflowing components, non-cirucular orbits and variability in the stellar properties could all smear out the correlation.

\subsection{The Double-Peak H${\alpha}$ Profile of PTF09djl}

\begin{figure}
\includegraphics[trim=0 0 0 0, clip, width=\columnwidth]{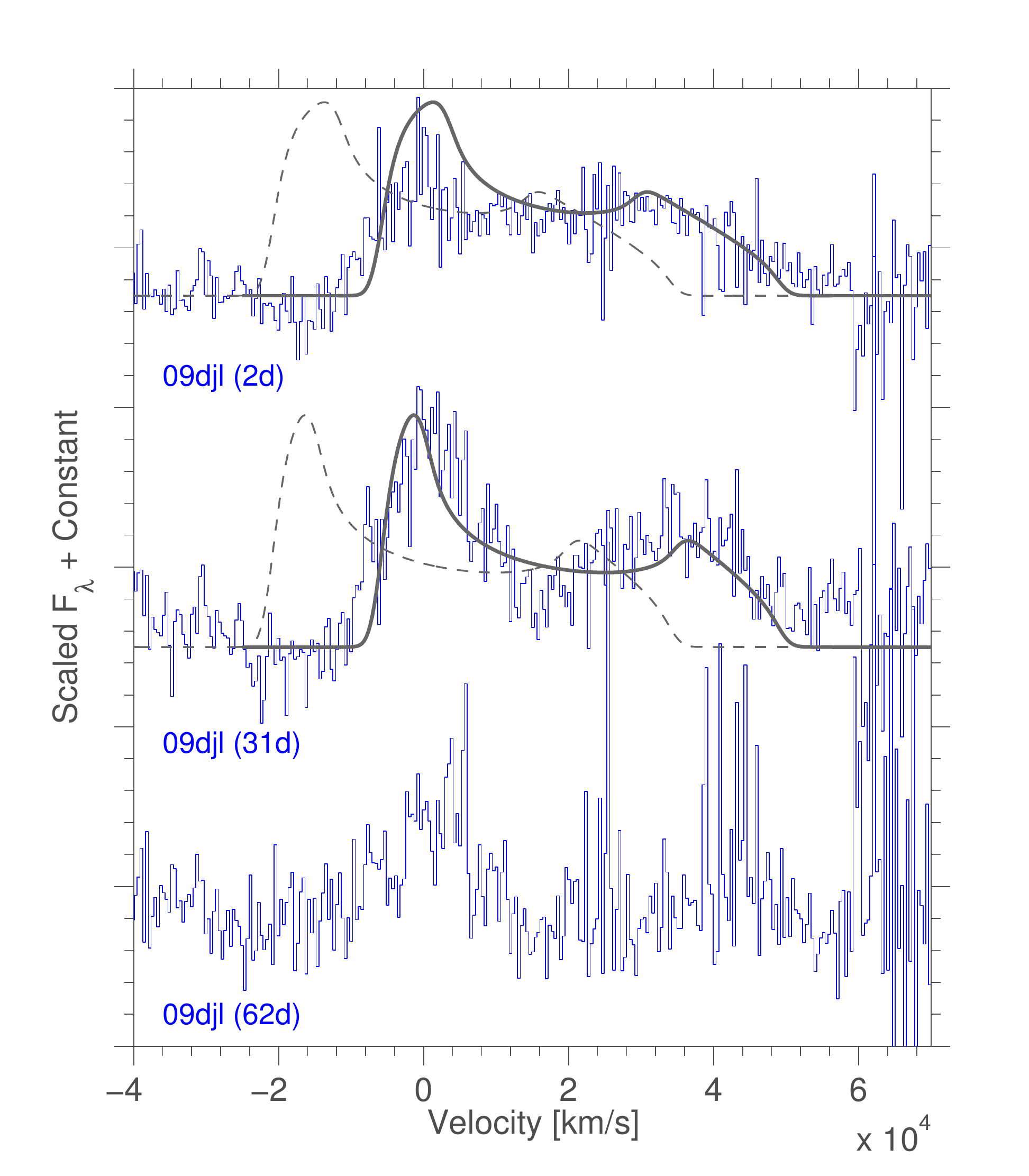}
\caption{\label{fig:djl_h}The H${\alpha}$ profile in the spectra of PTF09djl after host and continuum subtraction, corrected to the host redshift determined by its narrow absorption features. The data are binned to improve signal to noise, and phases are shown relative to peak. A red tail (first spectrum) and distinct red component (second spectrum) can be seen. A circular Keplerian disk model, following Strateva et al. (2003), is shown for inner and outer radii of $70$ and $300$ Schwarzschild radii (first spectrum) and $70$ and $170$ Schwarzschild radii (second spectrum), respectively. Both models assume an emissivity index of $-3$, a local turbulent broadening of $1200$ km~s$^{-1}$ and an inclination angle of $50$ degrees. The models are shifted to the red by $15000$ km~s$^{-1}$ (solid lines) from their original position (dashed lines).}
\end{figure}

For PTF09djl we find an additional redshifted component to the H${\alpha}$ emission feature extending out to high velocities  (Fig. \ref{fig:djl_h}). The observed structure is reminiscent of the double-peaked line profiles usually explained by Keplerian disk models (e.g. Chen et al. 1989). We construct a circular disk model following Strateva et al. (2003), which reproduces the shape but not the location of the profile (Fig. \ref{fig:djl_h}). While the model emission peaks are symmetric around the rest wavelength, in PTF09djl one peak is at the rest wavelength while the second is redshifted. We therefore have to shift the model profile to fit it to the observed spectrum. The disk model would thus have to include a bulk motion component to explain these observations.

Another possibility is that the geometry is more complex than a disk, with the emission related to the unbound debris, such as in the models considered by Bogdanovic et al. (2004). Their models reproduce the general profile shape but with much lower velocities (though the velocities can be increased using smaller inner-radii; Bogdanovic, private communication).

The double-peaked structure is not observed in the H${\beta}$ line nor is it seen in the other events (though this could be due to a low inclination angle for the disc model there).

\subsection{Radio Non-Detections of PTF09axc}

Our VLA observations of PTF09axc resulted in a null-detection at both the $3.5$\,GHz and $6.1$\,GHz with an RMS of $110\mu$Jy and $50\mu$Jy, respectively. At the redshift of PTF09axc these limits imply $3\sigma$ luminosity upper limits of $1.2\times 10^{29}$\,erg\,Hz$^{-1}$\,s$^{-1}$ and $5.3\times 10^{28}$\,erg\,Hz$^{-1}$\,s$^{-1}$ at these bands. 

So far, the only two TDE candidates ever to be detected in the radio are those found in $\gamma$-rays by {\it Swift}:  {\it Swift} J1644+57 (Bloom et al. 2011; Burrows et al. 2011; Levan et al. 2011; Zauderer et al. 2011) and  {\it Swift} J2058+05 (Cenko et al. 2012). The high energy emission of these events was suggested to originate in a relativistic jet pointed in our direction. The radio emission of such a jet should be observable also in off-axis cases.

\begin{figure}
\includegraphics[width=\columnwidth]{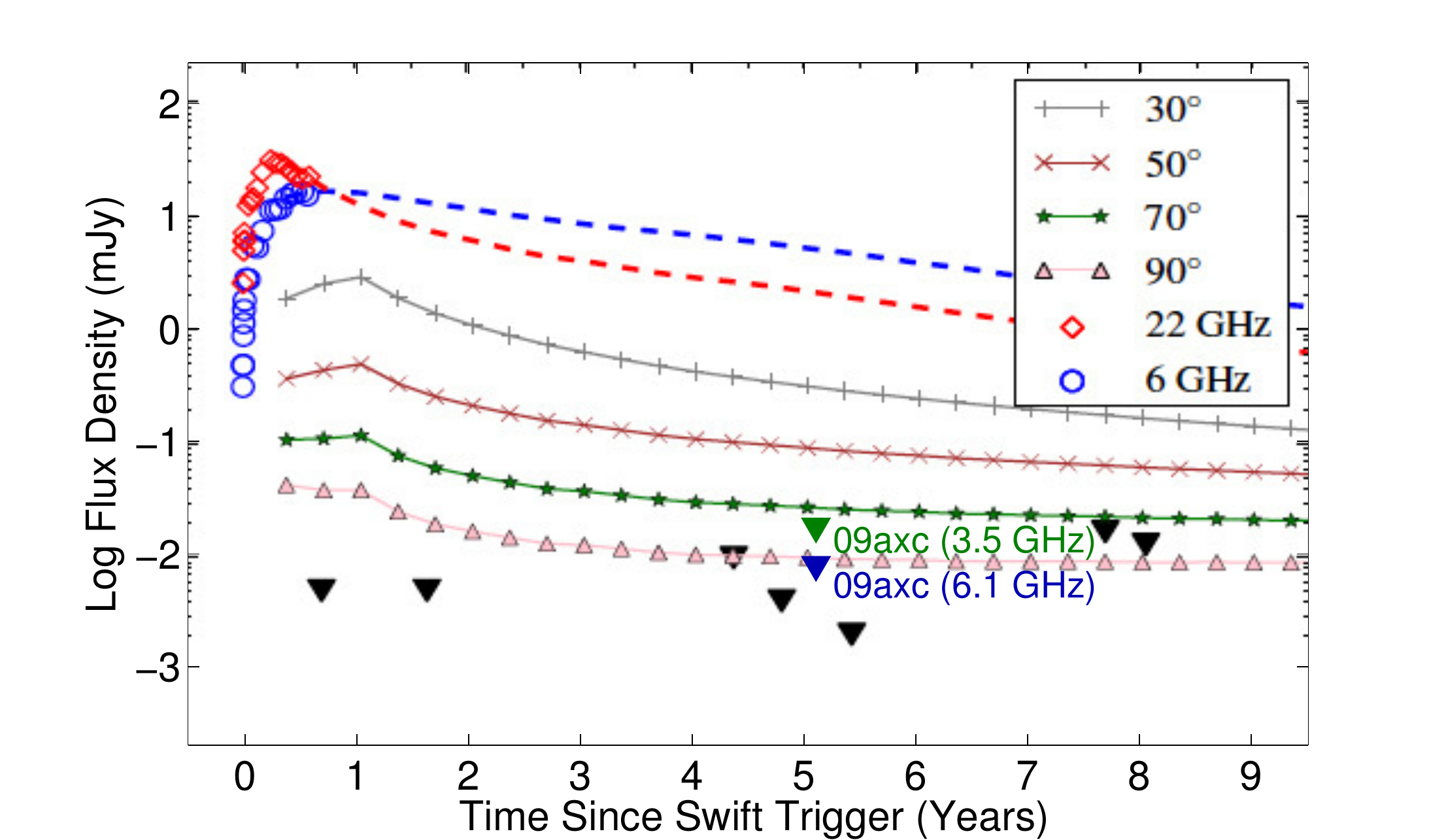}
\caption{\label{fig:radio} Radio non-detection limits of PTF09axc, $5$ years after the eruption, over-plotted on Figure 1 from van Velzen et al. (2013). The open symbols show the observed radio light curve of {\it Swift} J1644+57. The dashed lines are predicted light curves from Berger et al. (2012) assuming a jet energy of $10^{52}$\,erg. The van Velzen et al. (2013) $5$\,GHz off-axis light curve predictions are also shown. The triangles denote $2\sigma$ upper limits on the radio flux of the van Velzen et al. (2013) sample (black) and our data (green and blue). All data and models are scaled to the redshift of {\it Swift} J1644+57 ($z=0.354$). An off-axis jet is very unlikely for PTF09axc.}
\end{figure}

Recently, van Velzen et al. (2013) observed seven TDE candidates (including PS1-10jh and TDE2) in the radio, and did not detect emission for any of them. The non-detection limits were deep enough to rule out most off-axis jet angles for these events and led van Velzen et al. (2013) to conclude that it is not likely that all TDEs launch jets. Our limits for PTF09axc strengthen this conclusion (Figure \ref{fig:radio}).

\section{Summary and Conclusions}

We presented the results of a search for blue transients with peak magnitudes between those of SNe and SLSNe in the PTF non-interacting core-collapse SN sample. Of the six events found, we focus on PTF09ge, PTF09axc and PTF09djl, the three which are coincident with the centers of their host galaxies. These events, selected amongst core-collapse SN candidates only by their peak magnitude, color, centrality in their hosts and lack of narrow emission lines indicative of interaction-powered emission, turn out to be similar photometrically to the TDE candidate PS1-10jh (G12), and are all found to reside in E+A (or E+A-like) galaxies. 

The lack of obvious AGN features in the spectra of the outbursts and the host galaxies, as well as the magnitude of the outbursts and their non-recurrence, all disfavor an AGN interpretation for these events. The lack of recent or ongoing star formation in all the hosts rules out core-collapse SNe which originate in massive, short-lived stars. In addition, the blue spectra imply higher temperatures than those typically observed in SNe. We conclude that these events are likely TDEs.

One of the events is similar spectroscopically to the He-rich TDE candidate PS1-10jh, while the other two show broad hydrogen emission features, as expected by some models for TDE's involving a H-rich star. A spectrum of the recent TDE candidate ASASSN-14ae displays {\it both} H and He emission features. Comparing these events to literature data of two additional TDE candidates, we find a sequence of H- to He-dominated spectral features. This indicates that either a viewing-angle effect or some continuous parameter(s) of the disrupted star, its orbit, or the SMBH may generate a continuum of TDE spectral types. 

Van Velzen et al. (2011) calculate that PTF could potentially detect $\sim13$ TDE flares per year. However, there exists a strong bias against spectroscopically following transients in centers of galaxies by PTF or any transient survey which is wary of contamination by AGN and subtraction artifacts around galactic cores. Additional biases inherent to the survey (such as alternating observing strategies and followup prioritizations) make it impossible at this time to quantify the true expected rate of spectroscopically observed TDEs in PTF. 

The preference of these events for E+A hosts is intriguing. If indeed E+A's are post-merger galaxies (Zabludoff et al. 1996), then their TDE rate could be enhanced by the interaction of two central SMBHs. Chen et al. (2009) find that a binary SMBH can increase the TDE rate by several orders of magnitude, even up to $\sim1$ event per year (though only for a relatively short period). Some E+A galaxies may be the result of interaction rather than mergers (Goto 2005, Yamauchi et al. 2008). In this case, the TDE rate may be enhanced by dynamical interactions perturbing the orbits of stars near the SMBH. Alternatively, a unique stellar population in the cores of E+A galaxies may influence the TDE rate. A large fraction of evolved extended stars, for example, would increase the rate of TDEs since lower density stars are more easily disrupted (such stars may also experience multiple partial disruptions; MacLeod et al. 2013). We note, however, that our light curves are not as slowly evolving as those expected for the tidal disruptions of evolved stars (MacLeod et al. 2012). 

We note that when excluding SNe classified as Type IIn, we may have introduced a bias against TDEs in star-forming hosts (where the narrow emission lines from the host would have lead to the miss-classification of the transient as a SN IIn). However, the line profiles of IIn SNe are different than those of galaxies (e.g. Kiewe et al. 2012), and given that we use both visual and well-established SN spectral fitting tools to identify them, and that half the events which made the cut into our initial sample are in star-forming hosts, we conclude that any bias against such hosts is small, if at all present. Even if such a bias were strong, it would introduce an over-representation of passive galaxies in general, and not exclusively rare E+A's as we see here.
 
We detect X-ray flux at the position of PTF09axc, which is marginally consistent with an extremely weak AGN given our measured [O III] luminosity (Heckman et al. 2005) and highly inconsistent with an accreting binary origin (Hornschemeier et al. 2005). The high X-ray luminosity and low mass of the host galaxy are very similar to that of X171206.83+640830.7 (Hornschemeier et al. 2005). Curiously, this source may be variable and its host galaxy is also a post-merger (David et al. 2013). If some TDEs can emit X-rays at late time, it may be that X171206.83+640830.7 was also a TDE in an E+A galaxy.

One of the events in our PTF sample (PTF10iam) is clearly offset from the center of its host, and its light curve and spectral behavior are different than those of the central events. Its host galaxy is also different, showing signs of star-formation. We discuss this transient as a possible interacting SN (displaying broad high velocity H$\alpha$ absorption rather than the typical narrow emission lines) in a companion paper (Arcavi et al. {\it in prep}.).

The combined sample of nuclear transients presented here, tied together by the PTF events, now strongly supports a TDE origin for all of these objects, spanning a continuum of spectral classes, and preferring low star-forming E+A-like host galaxies. 

LSST will discover thousands of TDEs per year (van Velzen et al. 2011), but UV observations are required to accurately constrain the temperature and energetics of these transients. The proposed {\it ULTRASAT} mission could discover hundreds of TDEs per year in the UV (Sagiv et al. 2014). Understanding how to interpret TDE observations will enable them to be used to study accretion physics, stellar populations and otherwise-quiescent black holes.\newline

We thank R. Antonucci, L. Bildsten and C. S. Kochanek for helpful discussions. We appreciate the assistance of V. Bhalerao, A. Cucchiara, D. Levitan, A. Mishra, J. M. Silverman, R. Walters and O. Yaron in obtaining and reducing observations, and are grateful to the staffs at the various observatories where data were obtained. We thank the {\it Swift} PI N. Gehrels and the entire {\it Swift} team for authorizing, scheduling and carrying out our requested observations. 

This paper is based on observations obtained with the Samuel Oschin Telescope as part of the Palomar Transient Factory project. Additional data were obtained at the W. M. Keck Observatory, which is operated as a scientific partnership among the California Institute of Technology, the University of California and the National Aeronautics and Space Administration. The Observatory was made possible by the generous financial support of the W. M. Keck Foundation. This research used resources of the National Energy Research Scientific Computing Center, which is supported by the Office of Science of the U.S. Department of Energy under Contract No. DE-AC02-05CH11231. Part of this research was carried out at the Jet Propulsion Laboratory, California Institute of Technology, under a contract with the National Aeronautics and Space Administration. This research also made use of the NASA/IPAC Extragalactic Database (NED) which is operated by the Jet Propulsion Laboratory, California Institute of Technology, under contract with the National Aeronautics and Space Administration. The National Radio Astronomy Observatory is a facility of the National Science Foundation operated under cooperative agreement by Associated Universities, Inc. Funding for SDSS-III has been provided by the Alfred P. Sloan Foundation, the Participating Institutions, the National Science Foundation, and the U.S. Department of Energy Office of Science.

A.G. and I.A. acknowledge support by the Israeli Science Foundation and an EU/FP7/ERC grant. A.G. further acknowledges grants from the BSF, GIF and Minerva, as well as the ``Quantum Universe'' I-Core program of the planning and budgeting committee and the ISF, and a Kimmel Investigator award. E.O.O. is incumbent of the Arye Dissentshik career development chair and is grateful for support by a grant from the Israeli Ministry of Science and the I-CORE Program of the Planning and Budgeting Committee and The Israel Science Foundation (grant No 1829/12). M.M.K. acknowledges generous support from the Hubble Fellowship and Carnegie-Princeton Fellowship. J.S.B. and his group were partially supported by NASA/Swift Guest Investigator grants NNX09AQ66G and NNX10AF93G, and NSF/AST-100991. A.A.M. acknowledges support for this work by NASA from a Hubble Fellowship grant HST-HF-51325.01.

\end{document}